\documentclass[twocolumn,showpacs,preprintnumbers,amsmath,amssymb]{revtex4}
\usepackage{amsfonts}
\usepackage{graphicx}% Include figure files
\usepackage{dcolumn}% Align table columns him decimal of culminating points \usepackage{bm}% bold math

\begin{document}
\newcommand{\be}{\begin{equation}}
\newcommand{\ee}{\end{equation}}
\newcommand{\ber}{\begin{eqnarray}}
\newcommand{\eer}{\end{eqnarray}}
\newcommand{\mean}[1]{\left\langle #1 \right\rangle}
\newcommand{\abs}[1]{\left| #1 \right|}
\newcommand{\set}[1]{\left\{ #1 \right\}}
\newcommand{\la}{\langle}
\newcommand{\ra}{\rangle}
\newcommand{\lb}{\left(}
\newcommand{\rb}{\right)}
\newcommand{\norm}[1]{\left\|#1\right\|}
\newcommand{\RA}{\rightarrow}
\newcommand{\tet}{\vartheta}
\newcommand{\eps}{\varepsilon}
\newcommand{\tNN}{\tilde{\mathbf{X}}_n^{NN}}
\newcommand{\NN}{\mathbf{X}_n^{NN}}

\title{ Noise reduction in chaotic time series by a local projection
with nonlinear constraints}
\author{Krzysztof Urbanowicz}
 \email{urbanow@if.pw.edu.pl}
\affiliation{Faculty of Physics and Center of Excellence Complex
Systems Research \\Warsaw University of Technology \\
Koszykowa 75, PL--00-662 Warsaw, Poland}
\author{Janusz A. Ho{\l}yst}%
 \email{jholyst@if.pw.edu.pl}
\affiliation{Faculty of Physics and Center of Excellence Complex
Systems Research \\Warsaw University of Technology \\ Koszykowa
75, PL--00-662 Warsaw, Poland}
\author{Thomas Stemler}
\affiliation{Institute of Solid-State Physics\\Darmstadt
University of Technology\\Hochschulstr.6, D-64289 Darmstadt,
Germany}
\author{Hartmut Benner}
\affiliation{Institute of Solid-State Physics\\Darmstadt
University of Technology\\Hochschulstr.6, D-64289 Darmstadt,
Germany}
\date{\today}% It is always \today, today,

\begin{abstract}
On the basis of a local-projective (LP) approach we develop a
method of noise reduction in time series that makes use of
nonlinear constraints appearing due to the deterministic character
of the underlying dynamical system. The Delaunay triangulation
approach is used to find the optimal nearest neighboring points in
time series. The efficiency of our method is comparable to
standard LP methods  but our method is more robust to the input
parameter estimation.
 The approach has been successfully applied  for
separating a signal from noise in the chaotic Henon and Lorenz
models as well as for noisy experimental data obtained from an
electronic Chua circuit. The method works properly for a mixture
of additive and dynamical noise and can be used for the
noise-level detection.
\end{abstract}

 \maketitle

% ------------------------------------------------------------------------
\section{Introduction}
    \par It is common that observed data  are contaminated
 by noise (for a review of methods of
 nonlinear time series analysis see
 \cite{kantzschreiber,abarbanel,kapitaniak}). The presence of noise can substantially affect
 such system parameters as dimension,
     entropy or Lyapunov exponents \cite{urbanowicz}. In fact noise can completely obscure or even destroy the fractal
     structure of a chaotic attractor \cite{kostelich} and even $2\%$ of noise can make a dimension
     calculation misleading \cite{Schreiber1}. It follows that both from the theoretical as well as
     from the practical point of view it is desirable to reduce the noise level. Thanks to noise reduction
\cite{kostelich,Schreiber2,Farmer,hammel,davies,zhang,chen,Grassberger,kalman,effern,Hsu,Sauer,Schreiber}
      it is possible e.g. to restore the hidden
    structure of an attractor which is smeared out by noise,
    as well as to improve the quality of predictions.

    \par Every method of noise reduction assumes that it is possible to distinguish between noise
    and a {\it clean signal} on the basis of some objective criteria.
    Conventional methods such as linear filters use a power spectrum for this purpose.
    Low pass filters assume that a clean signal has some typical low frequency,
    respectively it is true for high pass filters.
    It follows that these methods are convenient for a regular source which generates a periodic or
    a quasi-periodic signal.
    In the case of chaotic signals linear filters cannot be used for noise reduction without a substantial disturbance of
    the clean signal. The reason is the broad-band spectrum of chaotic signals. It follows that for chaotic systems
    we make use of another generic feature of dissipative motion located on attractors that are
     smooth submanifolds of an admissible phase space.
    As results corresponding state vectors reconstructed
    from time delay variables are limited to geometric
    objects that can be locally linearized. This fact is a common background of all
    local projective (LP) methods of noise reduction.

    \par Besides the LP approach there are also noise reduction methods
    that approximate an unknown equation of motion and use it
    to find corrections to state vectors. Such methods make use
    of neural networks \cite{zhang} or a genetic programming \cite{chen} and one has to assume
    some basis functions e.g. radial basis functions
    \cite{broomhead} to reconstruct the equation of motion. Another group of
    methods are modified linear filters e.g. the Wiener
    filter \cite{Grassberger}, the Kalman filter \cite{kalman}, or methods based on wavelet analysis \cite{effern}.
    Applications of these methods are limited to systems with large sampling frequencies,
    and they are confined to the neighborhood of every point in phase space.

    \par The method described in this paper can be considered as an extension of LP
     methods by taking into account constraints that occur due to the local linearization of
    the equation of motion of the system. We call our method the
    \textit{local projection with nonlinear constraints} (LPNC).

\par The paper is organized as follows.
    In the following section we shall present the general background of LP methods.
    The LPNC method is introduced in Sec.~\ref{sec.1dimLPC} and compared with LP methods
    in Sec.~\ref{sec.comparison}.
    In Sec.~\ref{sec.findNN} we present methods how to find the nearest neighborhood, and examples of
    noise reduction and estimation are introduced in Secs.~\ref{sec.NRexample} and \ref{sec.estimation}.
    In the appendix~\ref{sec.multidimensional} one can find the multidimensional generalization
    of the solution presented in Sec.~\ref{sec.1dimLPC}.

  \section{\label{sec.LPmethods}Local Projective Methods of Noise Reduction}
\par Let us consider a scalar time series $\{\tilde{x}_n\}$, $n=1,2,...,N$ corresponding to
an experimentally accessible component of the system trajectory.
%$f: \tilde{\mathbf{x}}_{n+1}=f\lb\tilde{\mathbf{x}}_n\rb$.
We assume that in the presence of \textit{measurement noise}
instead of the clean time series $\tilde{x}_n$ we observe a noisy
series $x_n$: $x_n=\tilde{x}_n+\eta_n$ where $\eta_n$ is the noise
variable. The aim of noise reduction methods is to estimate the
set $\{\tilde{x}_n\}$ from the observed noisy data set $\{x_n\}$,
i.e. to find  corrections $\delta x_n$ such that $
x_n+\delta{x}_n\approx \tilde{x}_n$. The corrections $\delta x_n$
can be estimated on the assumption that $\tilde{x}_n$ belongs to a
clean deterministic trajectory. Let us create vectors of the
system state $\tilde{\mathbf{x}}_n$ using the Takens Theorem
\cite{abarbanel}
$\tilde{\mathbf{x}}_n=\{\tilde{x}_n,\tilde{x}_{n-\tau},...,\tilde{x}_{n-(d-1)\tau}\}$,
where $d$ is the embedding dimension, and $\tau$ is the embedding
delay that further will be just $1$. Now the simple approach is to
use a linear approximation for the nearest neighborhood
$\tilde{\mathbf{X}}_n^{NN}$ of a vector $\tilde{\mathbf{x}}_n$ and
then to estimate an unknown equation of motion in the embedded
space by a linear fit:
$\tilde{\mathbf{x}}_{n+1}=\mathbf{A}\tilde{\mathbf{x}}_n+\mathbf{b}$.
The matrix $\mathbf{A}$ is the corresponding Jacobi matrix and
$\mathbf{b}$ is a constant vector.

 In LP methods the local linearity of the system
dynamics plays the crucial role. The unknown equation of motion of
a deterministic systems $\tilde{x}_{n+1}=F\lb
\tilde{x}_n,\tilde{x}_{n-1},\ldots,\tilde{x}_{n-d+1}\rb$ is
equivalent to the presence of a constraint $H\lb
\tilde{x}_{n+1},\tilde{x}_n,\ldots,\tilde{x}_{n-d+1}\rb=0$. If the
embedding dimension $d$ is larger than the dimension of the
attractor then $Q$ constraints appear: \be H_q^{\lb n\rb}
\lb\tilde{x}_{n+1},\tilde{x}_n,\ldots,\tilde{x}_{n-d+1}\rb=0\quad\mbox{and}\quad
q=1,\ldots,Q\leq d \label{eq.wiazPL} \end{equation} where $Q$
depends on the rounded up dimension $d_a$ of the attractor,
$Q=d+1-d_a$. Since we apply a linear approximation for vectors
$\tilde{\mathbf{x}}_i\in \tNN$ the constraints (\ref{eq.wiazPL})
can be written as
\begin{eqnarray} H_q^{\lb n\rb}
\lb\tilde{x}_{n+1},\tilde{x}_n,\ldots,\tilde{x}_{n-d+1}\rb=\nonumber\\\sum\limits_{j=1}^{d-Q}
a_j^{q,\lb n\rb} \tilde{x}_{i-j+1-q}+b^{q,\lb n
\rb}-\tilde{x}_{i+1-q}=0,\label{eq.wiazPLroz}\end{eqnarray} where
$a_j^{q,\lb n\rb}$ and $b^{q,\lb n \rb}$ are elements of
$\mathbf{A}$ and $\mathbf{b}$ respectively.
 The main problem of LP methods is to find a tangent subspace
determined by the linear constraints $H_q^{\lb n
\rb}\lb\tilde{x}_{n+1},\tilde{x}_n,\ldots,\tilde{x}_{n-d+1}\rb=0$
and to perform an appropriate projection on this subspace.
Different LP approaches make use of different projecting methods,
however tangent subspaces are found in the same manner  by all
methods, i.e. the subspace should fulfill the condition
(\ref{eq.wiazPLroz}) and the condition
$\mean{\abs{x_i-\tilde{x}_i}^2}=min$.

\subsection{\label{sec.CHS} Cawley-Hsu-Sauer method (CHS)}
The method makes use of a perpendicular projection on a subspace
corresponding to the constraints (\ref{eq.wiazPLroz})
\cite{Hsu,Sauer}. Since there are several constraints
(\ref{eq.wiazPLroz}) and the same data will occur in several
Takens vectors $\mathbf{x}_n$ there are many possible corrections
$\delta x_{n,q}$ to the same observed data $x_n$. In the CHS
method one makes a compromise between different corrections by
taking the average
\begin{eqnarray} \tilde{\mathbf{x}}_n=\mathbf{x}_n+\alpha
\sum\limits_{q=1}^Q \delta \mathbf{x}_{n,q}\end{eqnarray} where
\begin{eqnarray}\delta \mathbf{x}_{n,q}=-H_q^{\lb n \rb}
\lb x_{n+1},x_n,\ldots,x_{n-d+1}\rb \frac{\bigtriangledown_n
H_q^{\lb n \rb}}{\norm{\bigtriangledown_n H_q^{\lb n
\rb}}^2}\end{eqnarray} is the correction of $\mathbf{x}_n$
obtained due to the constraint $H_q^{\lb n\rb} \lb
\tilde{x}_{n+1},\tilde{x}_n,\ldots,\tilde{x}_{n-d+1}\rb=0$.
$\alpha$ is some constant $0<\alpha<1$ and $\bigtriangledown_n
H_q^{\lb n \rb}=\bigtriangledown H_q^{\lb n\rb} \lb
x_{n+1},x_n,\ldots,x_{n-d+1}\rb$ is the gradient of the constraint
function.

\subsection{\label{sec.SG} Schreiber-Grassberger method (SG)}
 Instead of the perpendicular projection on the subspace defined
by (\ref{eq.wiazPLroz}) one can perform a projection by correcting
only one variable \cite{kostelich}.
 If we choose $x_{n+1-r}$ as the corrected variable where $r\approx d/2$ then the
corrections are \be \tilde{x}_n=x_n-\alpha \frac{H_s^{\lb n \rb
}\lb x_{n+1+r},x_{n+r},\ldots,x_{n-d+1+r}\rb}{\partial H_s^{\lb n
\rb }\lb x_{n+1+r},x_{n+r},\ldots,x_{n-d+1+r}\rb/\partial x_n}
\label{eq.SG}\end{equation} where $s\approx \frac{Q}{2}$. The
approach can be justified as follows. If the largest (unstable)
Lyapunov exponent is $\lambda_u
>0$ and the smallest (stable) Lyapunov
exponent is $\lambda_s<0$ we can write $\frac{\partial
x_{n+r}}{\partial x_n}\sim e^{\lambda_u r}$ and $\frac{\partial
x_{n-d+1+r}}{\partial x_n}\sim e^{\abs{\lambda_s} \lb r-d+1\rb}$.
If $\lambda_u\approx \abs{\lambda_s}$ then the highest precision
for determining the denominator of the rhs of (\ref{eq.SG}) is
usually obtained for $r=d/2$:
\begin{eqnarray} \sum\limits_{l=0}^{d/2}\abs{e^{\lambda_u
l}+e^{\abs{\lambda_s} l}} \abs{\Delta
x_n}=\nonumber\\\min\limits_{r=0,\ldots,d}\set{\sum\limits_{l=0}^r
\abs{e^{\lambda_u l}}\abs{\Delta x_n}+\sum\limits_{l=0}^{d-r}
\abs{e^{\abs{\lambda_s} l}}\abs{\Delta x_n}},\end{eqnarray}where
$\Delta x_n$ is the error connected with the variable $x_n$.

\subsection{\label{sub.Gpl}The optimal method of local projection (GHKSS)}
In the GHKSS  method \cite{Schreiber,kantzholyst} developed by
Grassberger \textit{et.al.} one looks for a minimization
functional that fulfills the linear constraints
(\ref{eq.wiazPLroz}) by corresponding corrections received in a
one-step procedure. The constraints (\ref{eq.wiazPLroz}) can be
written in the equivalent form $\lb\mathbf{a}^{q,\lb n \rb}\cdot
\tilde{\mathbf{y}}_n\rb+\mathbf{b}^{q,\lb n \rb}=0$ where  a new
vector
$\tilde{\mathbf{y}}_n=\set{\tilde{x}_{n+1},\tilde{x}_{n},\ldots,\tilde{x}_{n-d+1}}$
is introduced, the dimension of which is larger by one than the
dimension of the vector $\mathbf{x}_n$. Vectors $\mathbf{a}^{q,\lb
n \rb}$ should be linearly independent and appropriately
normalized, so that multiple corrections of the variables are
eliminated, i.e. $\mathbf{a}^{q,\lb n \rb}\cdot P\mathbf{a}^{q',
\lb n \rb}=\delta_{qq'}$ where $P$ is the matrix describing the
metric of the system. Let $\mathbf{Y}_n^{NN}$ be a set
corresponding to the nearest neighborhood of the vector
$\mathbf{y}_n$. Minimizing the functional $\sum\limits_k \delta
\mathbf{x}_k\cdot P^{-1}\delta \mathbf{x}_k$ for
$\set{k:\mathbf{y}_k\in\mathbf{Y}_n^{NN}}$ under the above
conditions we get a system of coupled equations. The next step is
to consider all vectors of $\mathbf{Y}_n^{NN}$ and to calculate
the average $\xi_i^{\lb n
\rb}=\frac{1}{\abs{\mathbf{Y}_n^{NN}}}\sum\limits_k x_{k+i}$,
$i=0,1,\ldots,d$ as well as corresponding $\lb d+1\rb\times\lb
d+1\rb$ covariance matrix  \be C_{ij}^{\lb n
\rb}=\frac{1}{\abs{\mathbf{Y}_n^{NN}}}\sum\limits_k x_{k+i}
x_{k+j}-\xi_i^{\lb n \rb}\xi_j^{\lb n \rb}, \end{equation} here
$\abs{\mathbf{Y}_n^{NN}}$ means the number of elements in the set.
Defining $R_i=\frac{1}{\sqrt{P_i}}$ and $\Gamma_{ij}^{\lb n
\rb}=R_i C_{ij}^{\lb n \rb}R_j$ one can find $Q$ orthonormal
eigenvectors of the matrix $\Gamma^{\lb n \rb}$ corresponding to
its smallest eigenvalues $\mathbf{e}^{q,\lb n \rb}$ for
$q=1,\ldots,Q$.

Let  the  matrix $\Pi_{ij}^{\lb n \rb}=\sum\limits_{q=1}^Q
e_i^{q,\lb n \rb}e_j^{q,\lb n \rb}$ define a subspace spanned by
the eigenvectors  $\mathbf{e}^{q,\lb n \rb}$. Now the  corrections
to the observed signal can be written as   follows \be \delta
x_{n+i}=\frac{1}{R_i}\sum\limits_{j=0}^d \Pi_{ij}^{\lb n \rb} R_j
\lb \xi_j^{\lb n \rb}-x_{n+j}\rb.
\label{eq.korektyGPL}\end{equation}

\par We see that the GHKSS method does not employ multiple corrections resulting
from constraints (\ref{eq.wiazPLroz}), but only performs a smaller
number of corrections following the multiple occurrence of the
same variable $x_n$ in various vectors $\mathbf{y}_i$ :
$x_n\in\mathbf{y}_i$.
\par The solution (\ref{eq.korektyGPL}) is a generalization of the CHS
and SG methods. The main difference between the CHS method and the
GHKSS method is in the subspace of projection. While a
perpendicular projection of points is used in the first case,
projection is on a tangent subspace defined by the matrix $P$ in
the second case. The matrix $P$ should be diagonal and such that
the first and the last component of the vector $\mathbf{y}_n$ have
only small weights e.g. :
  \be P = \begin{cases}
    0.001 & \text{i=0,d}, \\
    1 & \text{otherwise}.
  \end{cases}\end{equation}

%\be P_i=\set{
%  \begin{array}{cc}
%    0.001, & i=0,d \\
%    1, & i=1,\ldots,d-1
%  \end{array}}.\end{equation}

\par
The efficiency of noise reduction methods can be measured by the
gain parameter, defined as \be \mathcal{G}=10 \log\lb
\frac{\sigma_{noise}^2}{\sigma_{red}^2}\rb\end{equation} where
$\sigma_{noise}^2=\mean{\lb x_n-\tilde{x}_n\rb^2}$ is the variance
of added noise and $\sigma_{red}^2$ is the variance of noise left
after noise reduction. The last value is calculated as the square
of the distance between the vector of noise-reduced data and the
vector of clean data divided by the dimension of these vectors.
The definition of the gain presumes the knowledge of the clean
data $\tilde{X}_n=\set{\tilde{x}_n}$. \par The noise level
parameter $\mathcal{N}$ can be defined as the ratio of standard
noise deviation $\sigma_{noise}$ to standard data deviation
$\sigma_{data}$ \be
\mathcal{N}=\frac{\sigma_{noise}}{\sigma_{data}}.\label{eq.NTSdef}\end{equation}

\section{\label{sec.1dimLPC}The principle of LPNC method}
\par The LP methods described in the previous section make use of
linear constraints that appear due to linear approximation of the
system dynamics. Such a linear approximation has only a local
character and  corresponding coefficients depend, in fact, on the
position in phase space. If we assume that the nearest
neighborhood of every point $\mathbf{\tilde{x}}_n$ is
characterized by the same coefficients then nonlinear constraints
appear that can be used for reconstruction of the unknown
deterministic trajectory. The basic advantage of the \textit{local
projection with nonlinear constraints} (LPNC) method introduced
here as compared to LP methods is its smaller sensitivity to the
input parameters estimation. A weak point of the LPNC method is
its slower convergence rate with respect to the standard LP
approach. The LPNC algorithm can be accelerated but at the cost of
decreasing the gain parameter. Like other LP methods the LPNC
method belongs to the iterative approaches. A single iteration
provides only a partial noise reduction and a corrected data set
serves as an input for the next iteration.

%\par We present the idea of the LPNC method for the one-dimensional
%case. Generalization to a multi-dimensional problem is
%straightforward.
% and described in the
%Appendix~\ref{sec.multidimensional}.

\par For the one-dimensional case the Jacobi matrix $\mathbf{A}$ and
the additive vector $\mathbf{b}$ describing the locally linearized
dynamics at point $\tilde{x}_n$  reduce to scalar coefficients
$\mathbf{A}=a_1\lb\tilde{x}_n\rb$,
$\mathbf{b}=b\lb\tilde{x}_n\rb$, and the linearized equation of
motion at $\tilde{x}_n$ reads
$\tilde{x}_{n+1}=a_1\lb\tilde{x}_n\rb\tilde{x}_n+b\lb\tilde{x}_n\rb$.
 Let us consider the nearest neighborhood $\tilde{\mathbf{X}}_n^{NN}$ of $ \tilde{x}_n$.
 We assume that the set $\tilde{\mathbf{X}}_n^{NN}$
 consists of three points $\set{\tilde{x}_n,\tilde{x}_k,\tilde{x}_j\in\tNN}$ which are so {\it close} to each other
that their locally linearized  dynamics can be approximately
described by {\it the same} pair of coefficients
$\mathbf{A}=a_1\lb\tilde{x}_n\rb$,
$\mathbf{b}=b\lb\tilde{x}_n\rb$. When we write down three linear
equations of motion for
$\tilde{x}_{n},\tilde{x}_{k},\tilde{x}_{j}$ \ber
\tilde{x}_{n+1}=a_1\lb\tilde{x}_n\rb\tilde{x}_n+b\lb\tilde{x}_n\rb\nonumber\\
\tilde{x}_{k+1}=a_1\lb\tilde{x}_n\rb\tilde{x}_k+b\lb\tilde{x}_n\rb\nonumber\\
\tilde{x}_{j+1}=a_1\lb\tilde{x}_n\rb\tilde{x}_j+b\lb\tilde{x}_n\rb
\eer the coefficients $a_1(\tilde{x}_n)$ and $b(\tilde{x}_n)$ can
be eliminated. After elimination we get a constraint that has to
be fulfilled by the system variables for consistency reasons.

\begin{eqnarray} G\lb \tNN\rb\equiv
\tilde{x}_n\lb\tilde{x}_{k+1}-\tilde{x}_{j+1}\rb+\nonumber\\
\tilde{x}_k\lb\tilde{x}_{j+1}-\tilde{x}_{n+1}\rb+
\tilde{x}_j\lb\tilde{x}_{n+1}-\tilde{x}_{k+1}\rb=0.
\label{eq.wiaz1D}\end{eqnarray} In the case of a higher dimension
$d>1$ we have three equations of motions but the number of unknown
constants is larger than two, i.e.\ber
\tilde{x}_{n+1}=\sum_{i=1}^d a_i\lb\tilde{x}_n\rb\tilde{x}_{n-i+1}+b\lb\tilde{x}_n\rb\nonumber\\
\tilde{x}_{k+1}=\sum_{i=1}^d a_i\lb\tilde{x}_n\rb\tilde{x}_{k-i+1}+b\lb\tilde{x}_n\rb\nonumber\\
\tilde{x}_{j+1}=\sum_{i=1}^d
a_i\lb\tilde{x}_n\rb\tilde{x}_{j-i+1}+b\lb\tilde{x}_n\rb, \eer
where $a_{i}(\tilde{x}_n)$ are elements of the first row of Jacobi
matrix and $b(\tilde{x}_n)$ is a constant. The corresponding
constraint $G^d$ for higher dimensional case is as follows
\begin{eqnarray}
   G^d \lb \tNN\rb \equiv \lb\sum_{i=1}^d a_i x_{n-i+1}\rb\lb\tilde{x}_{k+1}-
   \tilde{x}_{j+1}\rb+\nonumber\\\lb\sum_{i=1}^d a_i x_{k-i+1}\rb\lb\tilde{x}_{j+1}-\tilde{x}_{n+1}\rb+\nonumber\\
\lb\sum_{i=1}^d a_i
x_{j-i+1}\rb\lb\tilde{x}_{n+1}-\tilde{x}_{k+1}\rb=0\label{eq.wiaz2D}
\end{eqnarray}
The extended constraints and the corresponding calculations that
are valid for all rows of Jacobi matrix $\mathbf{A}$ are presented
in the appendix~\ref{sec.multidimensional}.
 The condition~(\ref{eq.wiaz1D}) and~(\ref{eq.wiaz2D}) should be
fulfilled for every point $\tilde{x}_n$ and its nearest
neighborhood $\tNN$. Similarly as in LP methods these constraints
are ensured in the LPNC approach by application of the method of
Lagrange multipliers to an appropriate cost function. Since we
expect that corrections to noisy data should be as small as
possible, the cost function can be assumed to be the sum of
squared corrections $S=\sum_{s=1}^N \lb\delta x_s\rb^2$.
\par It follows that we are looking for the minimum of the functional \be
\tilde{S}=\sum_{n=1}^N\lb \delta x_n\rb^2+\sum_{n=1}^N\lambda_n
G^d\lb\tNN\rb= min. \label{eq.funkcjonal1}\end{equation} After
finding zero points of $2N$ partial derivatives one gets $2N$
equations with $2N$ unknown variables $\delta x_n$ and
$\lambda_n$. However, in such a case the derivatives of the
functional (\ref{eq.funkcjonal1}) are nonlinear functions of these
variables. For simplicity of computing we are interested to pose
our problem in such a way that linear equations appear which can
be solved by standard matrix algebra. To understand the role of
nonlinearity let us write the constraint $G\lb \tNN\rb$ in such a
way that explicit dependence on the unknown variables is seen (the
corresponding equations for $G^d(\tNN)$ have a similar form)
\begin{eqnarray} G\lb\tNN\rb\cong G\lb
\NN,\mathbf{X}_{n+1}\rb+G\lb \delta \mathbf{X}_n
,\mathbf{X}_{n+1}\rb+\nonumber\\G\lb
\NN,\delta\mathbf{X}_{n+1}\rb+G\lb \delta \mathbf{X}_n
,\delta\mathbf{X}_{n+1}\rb. \label{eq.rozwin}\end{eqnarray} Here
we introduced the following notation
\begin{widetext} \begin{eqnarray} G\lb \NN,\mathbf{X}_{n+1}\rb\equiv x_n\lb
x_{k+1}-x_{j+1}\rb+x_k\lb x_{j+1}-x_{n+1}\rb+x_j\lb
x_{n+1}-x_{k+1}\rb\nonumber\\
G\lb \delta \mathbf{X}_n ,\mathbf{X}_{n+1}\rb\equiv \delta x_n\lb
x_{k+1}-x_{j+1}\rb+\delta x_k\lb x_{j+1}-x_{n+1}\rb+\delta x_j\lb
x_{n+1}-x_{k+1}\rb\nonumber\\
G\lb \NN,\delta\mathbf{X}_{n+1}\rb\equiv x_n\lb \delta x_{k+1}-
\delta x_{j+1}\rb+x_k\lb  \delta x_{j+1}- \delta x_{n+1}\rb+x_j\lb
 \delta x_{n+1}- \delta x_{k+1}\rb\nonumber\\
G\lb \delta \mathbf{X}_n ,\delta\mathbf{X}_{n+1}\rb\equiv \delta
x_n\lb \delta x_{k+1}- \delta x_{j+1}\rb+\delta x_k\lb  \delta
x_{j+1}- \delta x_{n+1}\rb+\delta x_j\lb
 \delta x_{n+1}- \delta x_{k+1}\rb,\label{eq.expandg}\end{eqnarray} \end{widetext}

 where
 $\NN=\set{x_n,x_k,x_j}$, $\mathbf{X}_{n+1}=\set{x_{n+1},x_{k+1},x_{j+1}}$,
 $\delta \mathbf{X}_n=\set{\delta x_n,\delta x_k,\delta x_j}$,
  $\delta \mathbf{X}_{n+1}=\set{\delta x_{n+1},\delta x_{k+1},\delta x_{j+1}}$
 and $x_k,x_j$ are the near neighbors of $x_n$. Indices are defined as
 $\set{n,j,k:x_n,x_k,x_j\in\NN}$. Note that elements of the set $\mathbf{X}_{n+1}$
are not necessarily near neighbors to each other.
\par
 The approximation we use in (\ref{eq.rozwin}) follows from the fact that in general the
 nearest neighborhood $\tNN$ does not include the same indices as the nearest neighborhood
  $\NN$, i.e. \be
\set{k:\tilde{x}_k\in
 \tNN}\neq\set{j:x_j\in
 \NN}.\end{equation}
In the case of not correlated noise and under the assumption that
the introduced corrections completely reduce the noise effect
$\delta x_s=-\eta_s\quad(\forall_{s=1,\ldots,N})$ one can neglect
the nonlinear terms in Eqs.~(\ref{eq.expandg}) i.e.
\begin{equation}
 G\lb \delta \mathbf{X}_n,\delta
 \mathbf{X}_{n+1} \rb \cong 0\quad (\forall_{n=1,\ldots,N}) \label{eq.przybliz}.
\end{equation}
In the equation~(\ref{eq.przybliz}) we use the fact that
$\mean{\eta_i}=0$ and $\mean{\eta_i \eta_j} \sim \delta_{ij}$.
 \par
    Taking into account the approximation~(\ref{eq.przybliz}) one can write
    the following \textit{linear equation} for the problem~(\ref{eq.funkcjonal1})
    \be
    \mathbf{M} \cdot \delta
    \mathbf{X}=\mathbf{B},\label{eq.liniowef}
    \end{equation}
    where $\mathbf{M}$ is a matrix containing constant elements,
    $\mathbf{B}$ is a constant vector, and $\delta
    \mathbf{X}^T=\set{\delta x_1,\delta x_2,\ldots,\delta x_N,
    \lambda_1,\lambda_2,\ldots,\lambda_N}$ is a vector of dependent
    variables ($T$ - transposition). In practice it is very difficult or even impossible
    to find the solution of the equation~(\ref{eq.liniowef}) for large N.
    First,it is time consuming to solve a linear equation with a matrix $2N\times 2N$ matrix for $N>1000$.
    Second, when $\mathbf{M}$ becomes singular the estimation error of the inverse matrix $M^{-1}$ is very large.
    Third, we cannot always find the true near neighbors (the set $\tNN$) from
    the noisy data $\set{x_i}$. Taking into account the above reasons it is useful to replace
     the global minimization problem (\ref{eq.funkcjonal1}) by $N$ local minimization problems
     related to the nearest neighborhood
     $\NN $. The corresponding local functionals to be minimized are\begin{eqnarray}
     \tilde {S}_n^{NN}=\sum_s \lb \delta x_s
    \rb^2+\lambda_n G^d \lb \NN\rb = min\nonumber\\ (\forall_{n=1,...,N})\quad\mbox{where} \nonumber\\
    \quad\set{s:x_s\in \NN\;\mbox{or}\;x_s\in
    \mathbf{X}_{n+1}\;\mbox{or}\;\ldots\;x_s\in
    \mathbf{X}_{n-d+1}}.\label{eq.minimalizlok}
    \end{eqnarray}
    We can consider the minimization problem~(\ref{eq.minimalizlok}) as a certain
approximation of (\ref{eq.funkcjonal1}).
Functionals~(\ref{eq.minimalizlok}) are linked to each other due
to the fact that the same variable $\delta x_n$ appears in $6\cdot
d$ different minimization problems~(\ref{eq.minimalizlok}). The
global problem~(\ref{eq.funkcjonal1}) is equivalent to
Eq.~(\ref{eq.liniowef}) with $2N$ unknown variables that should be
found single-time. The problem (\ref{eq.minimalizlok}) is
equivalent to a system of coupled equations that should be solved
several times and as a result one gets an approximate global
solution. Writing Eq.~(\ref{eq.minimalizlok}) in the linear form
i.e. calculating zero sites of corresponding derivatives and using
Eq.~(\ref{eq.przybliz}) one gets $N$ linear equations as follows
\be  \mathbf{M}_n \cdot \delta
\mathbf{X}_n^{\lambda}=\mathbf{B}_n\quad(\forall_{n=1,\ldots,N}),\label{eq.liniowezlok}\end{equation}
where $\lb\delta \mathbf{X}_n^{\lambda}\rb^T=\set{\delta
x_n,\delta x_k,\delta x_j,\delta x_{n+1},\delta x_{k+1},\delta
x_{j+1},\lambda_n}$. The matrices $\mathbf{M}_n$ corresponding to
(\ref{eq.minimalizlok}) avoid the disadvantages of
(\ref{eq.liniowef}), i.e. they are not singular, their dimension
is smaller and they do not substantially depend on the initial
approximation of near neighbors. The matrix $\mathbf{M}_n$ for
one-dimensional case is given by
\begin{widetext}
\begin{eqnarray} \mathbf{M}_n=\left[
  \begin{array}{ccccccc}
    2 & 0 & 0 & 0 & 0 & 0 & x_{k+1}-x_{j+1}\\
    0 & 2 & 0 & 0 & 0 & 0 & x_{j+1}-x_{n+1}\\
    0 & 0 & 2 & 0 & 0 & 0 & x_{n+1}-x_{k+1}\\
    0 & 0 & 0 & 2 & 0 & 0 & x_j-x_k\\
    0 & 0 & 0 & 0 & 2 & 0 & x_n-x_j\\
    0 & 0 & 0 & 0 & 0 & 2 & x_k-x_n\\
    x_{k+1}-x_{j+1} & x_{j+1}-x_{n+1} & x_{n+1}-x_{k+1} & x_j-x_k & x_n-x_j & x_k-x_n & 0
  \end{array}
\right]
 \end{eqnarray}\end{widetext}
   Vector $\mathbf{B}_n$ has the form
    $\mathbf{B}_n^T=\set{0,0,0,0,0,0,-G\lb\NN,\mathbf{X}_{n+1}\rb}$.

\section{\label{sec.comparison}Comparing LPNC method to local projection methods}
    Let us illustrate the LPNC method by taking into account the cost functional~(\ref{eq.minimalizlok})
    (it will be written as $S^{LPNC}$)
%\begin{widetext}
\begin{eqnarray}
 S^{LPNC}=\sum_i \delta x_i^2 +\lambda [ \tilde{x}_n\lb \tilde{x}_{k+1}-\tilde{x}_{j+1}\rb+\nonumber\\
 \tilde{x}_k\lb \tilde{x}_{j+1}-\tilde{x}_{n+1}\rb+\tilde{x}_j\lb
 \tilde{x}_{n+1}-\tilde{x}_{k+1}\rb ]= \min. \label{eq.SPLzW}
\end{eqnarray}
%\end{widetext}
The corresponding cost function $S^{GHKSS}$ that is used in the
standard local projection method e.g. in the GHKSS method
\cite{Schreiber} is
%\begin{widetext}
\begin{eqnarray}
 S^{GHKSS}=\sum_i \delta x_i^2 + \lambda_1 \lb \tilde{x}_n a+b-\tilde{x}_{n+1}\rb+\nonumber\\
  \lambda_2 \lb \tilde{x}_j a+b-\tilde{x}_{j+1}
 \rb+ \lambda_3 \lb \tilde{x}_k
 a+b-\tilde{x}_{k+1}\rb= \min. \label{eq.SGPL}
\end{eqnarray}
%\end{widetext}
If we were in the position to find exact solutions for the
minimization problems (\ref{eq.SPLzW}) and (\ref{eq.SGPL}) then
both results would be the same since (\ref{eq.SPLzW}) can be
obtained from (\ref{eq.SGPL}) after elimination of the parameters
$a$ and $b$.
\par In both cases the variables
$\left\{\tilde{x}_n,\tilde{x}_k,\tilde{x}_j\in
\tilde{\mathbf{X}}_n^{NN}\right\}$ belong to the nearest
neighborhood of the variable $\tilde{x}_n$. The index
$i=\left\{k,k+p:\tilde{x}_k\in \tilde{\mathbf{X}}_n^{NN}\right\}$
runs through all indices of the variables appearing in
(\ref{eq.SPLzW}) and (\ref{eq.SGPL}) while the variable
$\tilde{x}_{k+p}=\set{\tilde{x}_l :
\tilde{x}_{l-p}\in\mathbf{\tilde{X}}_n^{NN}}$ corresponds to the
$p$-iterate of $\tilde{x}_k$. Parameters $a$ and $b$ can be
calculated from a linearized form of the equations of motion at
the point $\tilde{x}_n$.

\par In practice the minimization problems $S^{LPNC}$ and $S^{GHKSS}$ are not equivalent
 because in both cases different approximations are used. These differences
 are:
 (i) Eq.~(\ref{eq.SPLzW}) is nonlinear against corrections $\delta x_i$. In this case
 the approximation consists in a linearization.
 (ii) For Eq.~(\ref{eq.SGPL}) the exact values of the parameters $a$ and $b$ are unknown.
  The approximation means that $a$ and $b$ are estimated from noisy data.

\begin{figure}
\includegraphics[angle=-90,scale=0.35]{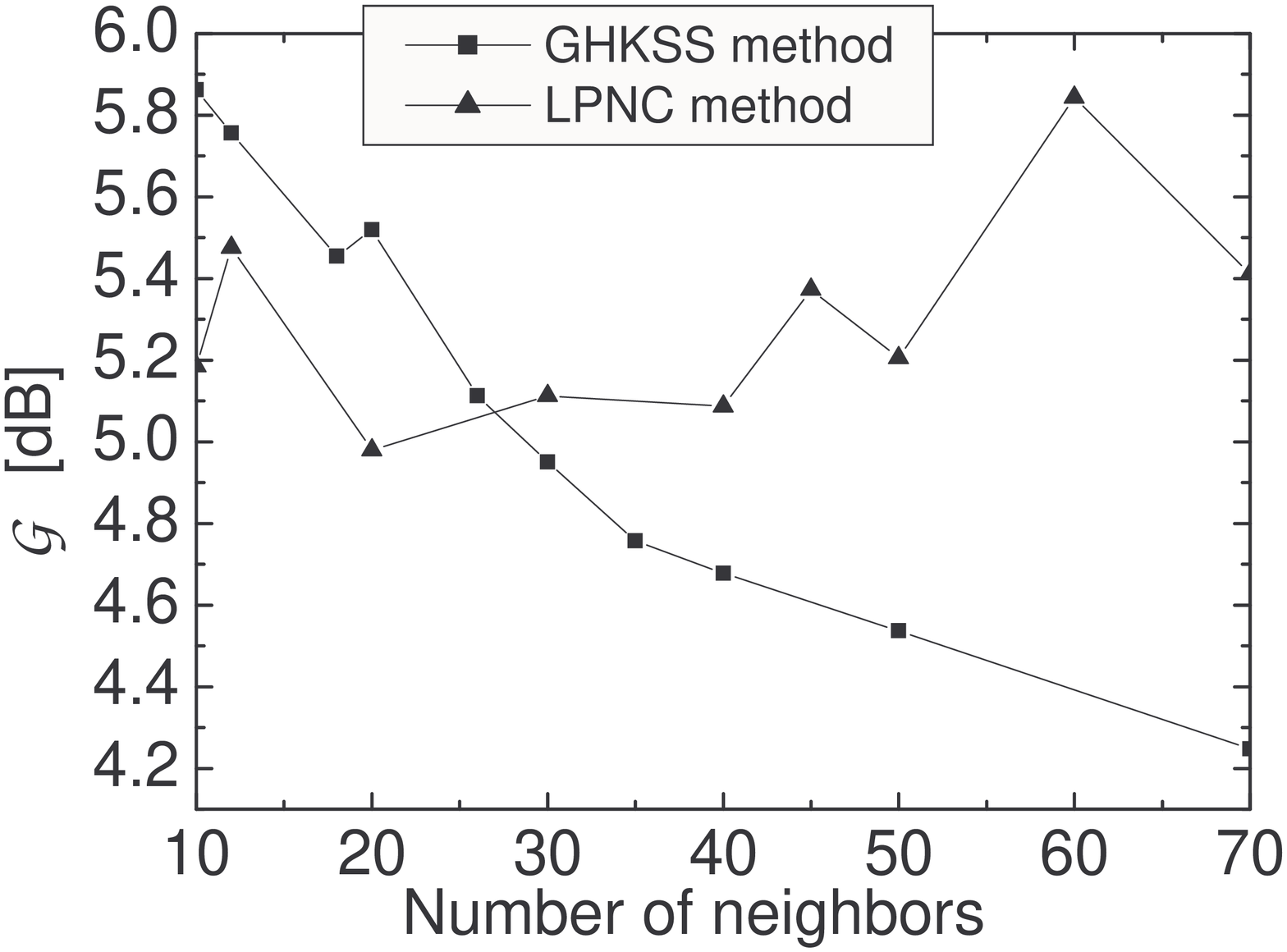}
\caption{\label{fig.nnGHKSSLPNC} The plot of the gain parameter
$\mathcal{G}$ versus number of iterations of the GHKSS method
(squares) and LPNC method (triangles). Lorenz system
$\mathcal{N}=78\%$, $N=1000$.}
\end{figure}

\begin{figure}
\includegraphics[angle=-90,scale=0.35]{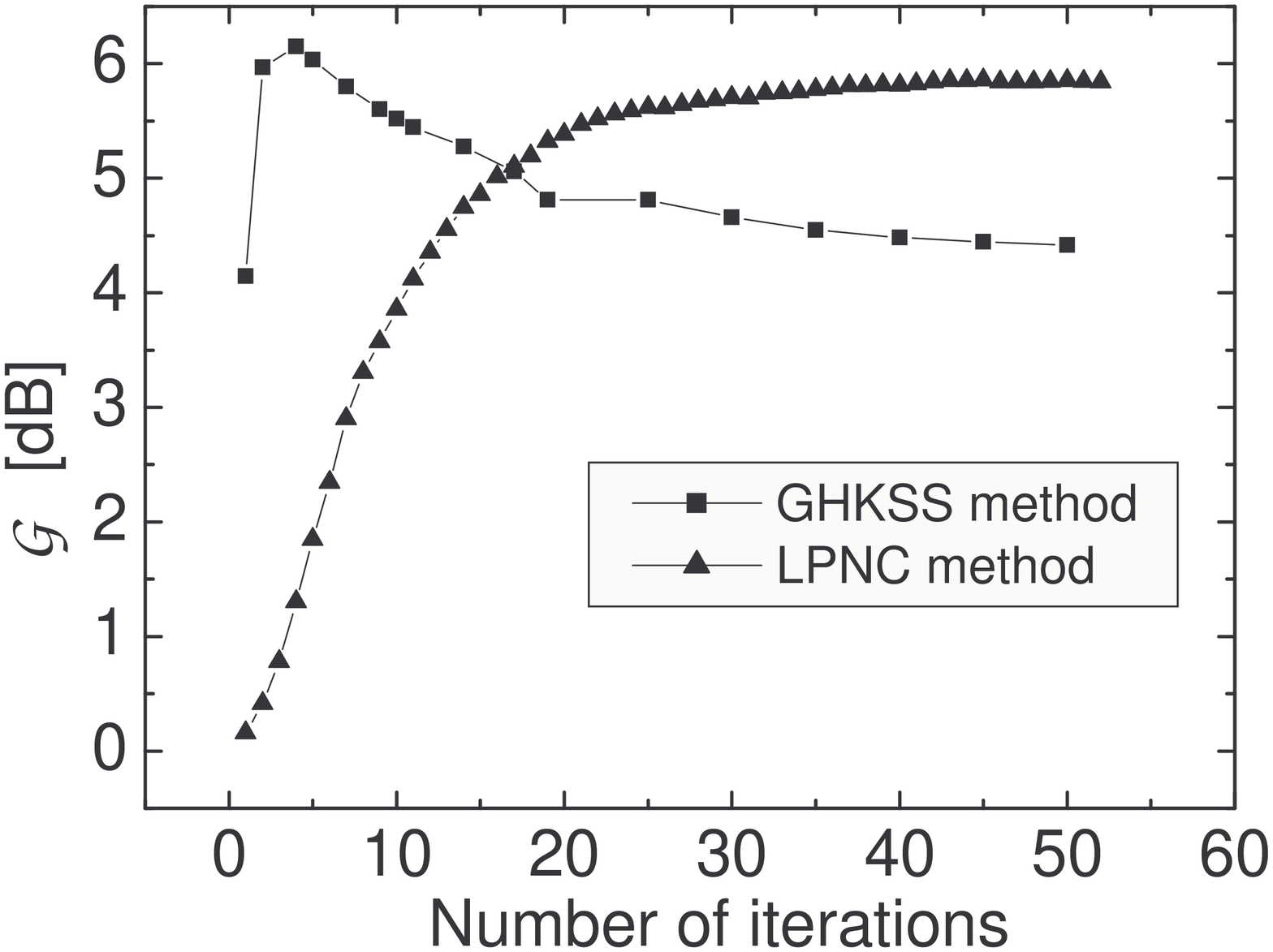}
\caption{\label{fig.iterGHKSSLPNC} The plot of the gain parameter
$\mathcal{G}$ versus number of neighbors of the GHKSS method
(squares) and LPNC method (triangles). Lorenz system
$\mathcal{N}=78\%$, $N=1000$.}
\end{figure}

Fig.~\ref{fig.nnGHKSSLPNC} and~\ref{fig.iterGHKSSLPNC} present a
comparison between results received by the GHKSS and LPNC methods.
Fig.~\ref{fig.nnGHKSSLPNC} shows that the gain parameter
$\mathcal{G}$ depends on the number of neighbors, which is an
input parameter of both methods. One can see that for LPNC method
the gain parameter is more robust to changes of the number of
neighbors than for the GHKSS method. In
Fig.~\ref{fig.iterGHKSSLPNC} the dependence of the gain parameter
on the number of iteration steps of the methods is shown. One can
see that LPNC method finished reduction at the maximal efficiency
what is not the case of GHKSS method, so the former method is
easier to use since it does not need estimation of the iteration
number.

\par
\begin{figure}
\includegraphics[angle=-90,scale=0.35]{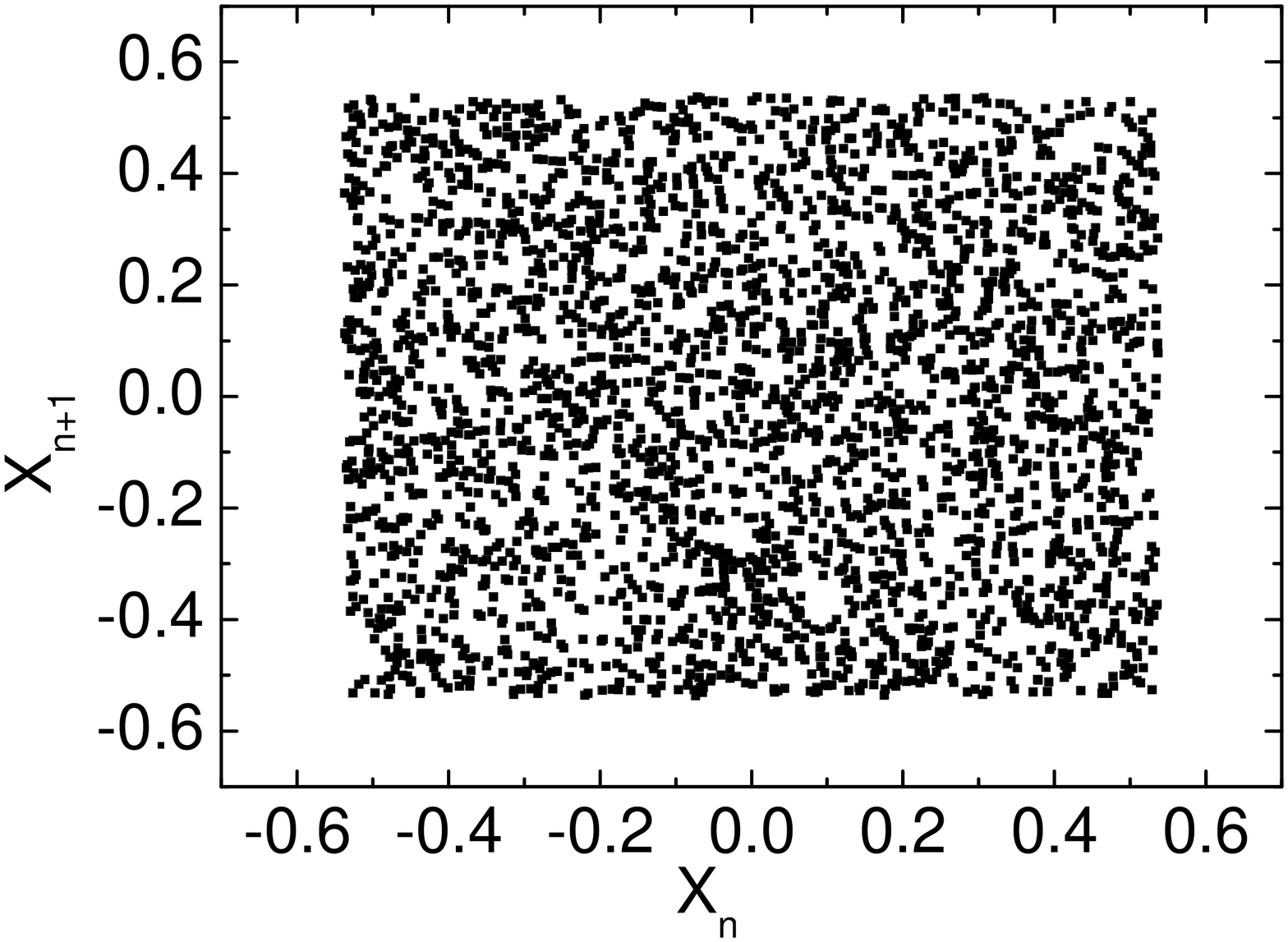}
\caption{\label{fig.szuma} The random data from uniform
distribution}
\end{figure}

\begin{figure}
\includegraphics[angle=-90,scale=0.35]{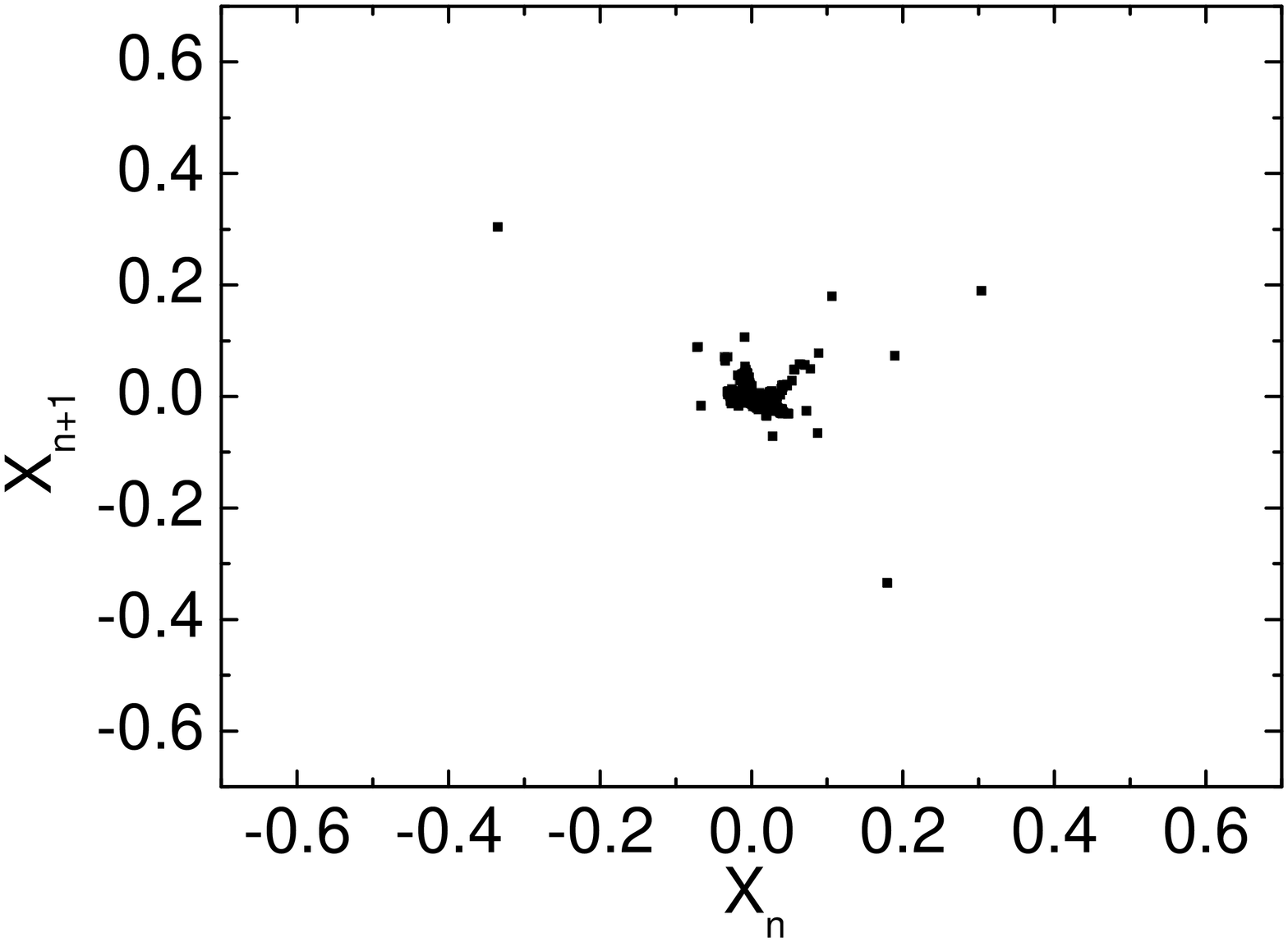}
\caption{\label{fig.szumc} The random data shown in the
Fig.~\ref{fig.szuma} after noise reduction with the LPNC method}
\end{figure}

 If we consider uniformly distributed stochastic
variables (see Fig.~\ref{fig.szuma}) the LPNC method reduces the
noise very well, and as a result all data are represented as a
neighborhood of a point attractor (see Fig.~\ref{fig.szumc}) while
a complete noise reduction would correspond to a phase portrait
consisting of a single point. In fact, for the case considered we
observed for the LPNC method a noise reduction of about $96\%$ of
data variance.

\section{\label{sec.findNN}The Nearest neighborhood assessment}
\par The LP methods are local.
It follows that features of the nearest neighborhood $\NN$ of
every point $\mathbf{x}_n$ in the phase space play an important
role.  Usually the nearest neighborhood is estimated by the
smallest distance approach that makes use of the standard
Euclidian geometry. We have found, however, that our LPNC method
works much better when the Delaunay triangulation approach
\cite{allie} is applied
 for the nearest neighborhood estimation.
\subsection{\label{sub.SD} The smallest distance approach (SD)}
\par In the smallest distance approach the Euclidian metric is used, i.e.
first the distance between every pair of points  in the Takens
embedded space is calculated as $d_{i,j}=\sqrt{\lb
x_i-x_j\rb^2+\ldots+\lb x_{i-\lb d-1\rb\tau}-x_{j-\lb
d-1\rb\tau}\rb^2}$ and then  the nearest neighborhood $\NN$ of a
point $\mathbf{x}_n$ is defined as a set of $\nu$ points
fulfilling the relation \be \set{\mathbf{x}_j\in\NN,\forall_k
\mathbf{x}_k\notin \NN:d_{n,j}\leq d_{n,k}}.
\label{eq.NONNdef}\end{equation} Let us stress that this
definition depends on the chosen value of the $\nu$ parameter.
i.e. on the assumed number of near neighbors, $\nu=2,3,\ldots$.
\subsection{\label{sub.Td} The Delaunay triangulation approach (DT)}
\par
\begin{figure}
\includegraphics[scale=0.9]{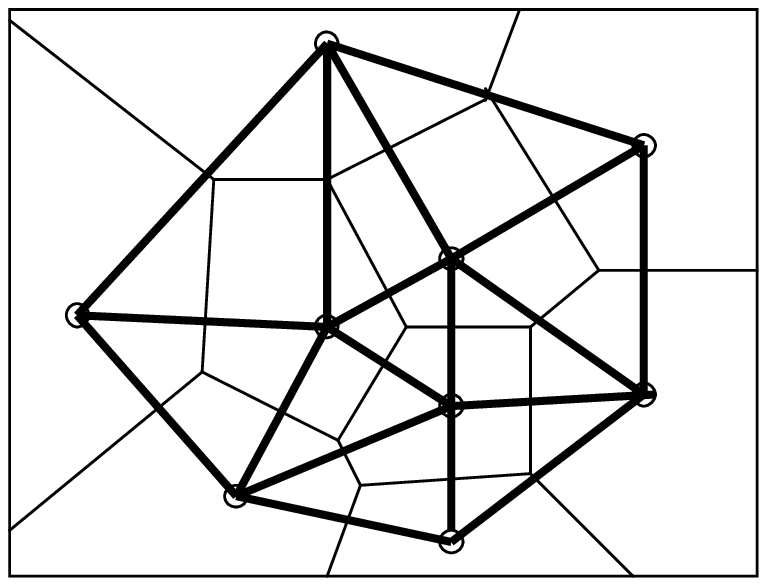}
\caption{\label{fig.TD} Delaunay triangulation for a set of nine
points in a two-dimensional space. The near neighbors are
connected by bold lines. Sets $T\lb \mathbf{x}_i\rb$,
$i=1,2,\ldots,9$ are limited by thin lines.}
\end{figure}

 To find the nearest neighborhood relations for the LPNC method
we have used  the Delaunay triangulation \cite{allie}. In general
the triangulation of any set of points
$\mathbf{X}=\set{\mathbf{x}_i}\in \mathbb{R}^d$ is a collection of
$d$-dimensional simplices with disjoint interiors and vertices
chosen from $\mathbf{X}$. There are many triangulation of the same
set of points $\mathbf{X}$. One of the best known is the Delaunay
triangulation (see Fig.~\ref{fig.TD}). Let $T\lb \mathbf{x}_n\rb$
be a part of the space $\mathbb{R}^d$ that contains all points
that are closer to $\mathbf{x}_n$ than any other point
$\mathbf{x}_j$ from the set $\mathbf{X}$ \be T\lb
\mathbf{x}_n\rb=\set{z\in
\mathbb{R}^d,\mathbf{x}_j\in\mathbf{X}:\forall_{j\neq
n}\norm{z-\mathbf{x}_n}\leq\norm{z-\mathbf{x}_j}}.\end{equation}
If $\mathbf{m}_{ i,j}=\lb\mathbf{x}_i+\mathbf{x}_j\rb/2$ belongs
to both sets $T\lb \mathbf{x}_i\rb$ and $T\lb \mathbf{x}_j\rb$
then by definition the point $\mathbf{x}_j$ is the nearest
neighbor of $\mathbf{x}_i$ received due to the Delaunay
triangulation.  By the above definition every point $\mathbf{x}_n$
belongs to its nearest neighborhood $\mathbf{x}_n\in \NN$.

\par In practice the Delaunay approach can be performed as follows.
A pair of points $\mathbf{x}_n$ and $\mathbf{x}_j$ are near
neighbors provided that there are no other points $\mathbf{x}_k$
($k\neq j,n$) belonging to the hypersphere centered at the point
$\mathbf{m}_{n,j}$ and of the radius
$r_{n,j}=\norm{\mathbf{x}_n-\mathbf{x}_{j}}/2$.

\begin{figure}
\includegraphics[scale=0.65]{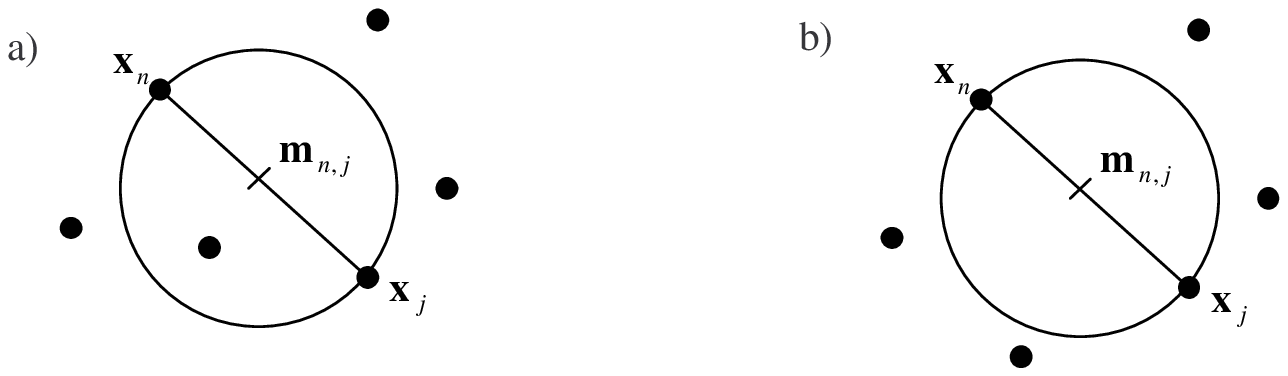}
\caption{\label{fig.TDprzypadki} Illustration of nearest
neighborhood search by DT approach a) $\mathbf{x}_j$ and
$\mathbf{x}_n$ are not near neighbors. b) $\mathbf{x}_j$ and
$\mathbf{x}_n$ are near neighbors}.
\end{figure}

In Fig.~\ref{fig.TDprzypadki} two cases are presented when in the
two-dimensional space a) the point $\mathbf{x}_j$ is not the
nearest neighbor of $\mathbf{x}_n$ and b) the point $\mathbf{x}_j$
is the nearest neighbor of $\mathbf{x}_n$.

\par The DT method has the advantage that triangles appearing due
to connections of near neighbors are almost equiangular (see
Fig.~\ref{fig.TD}). This property is the main reason for using the
DT method in search of the near neighbors. The disadvantage of
this method is a slowing down of numerical calculations.

\section{\label{sec.NRexample}Examples of noise reductions}
\par
\begin{figure}
\includegraphics[angle=-90,scale=0.35]{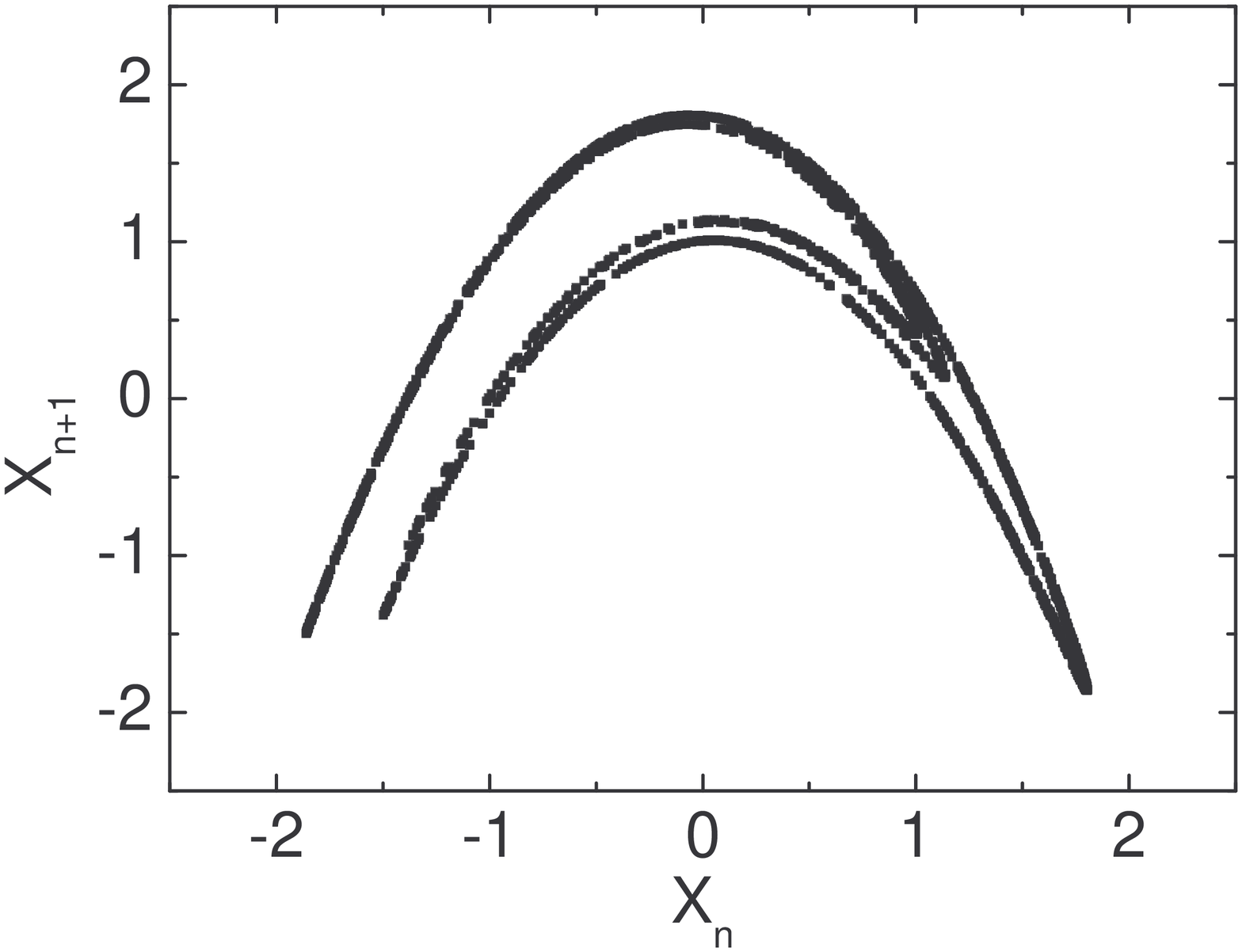}
\caption{\label{fig.henon100a} Chaotic Henon map without noise}
\end{figure}
\begin{figure}
\includegraphics[angle=-90,scale=0.35]{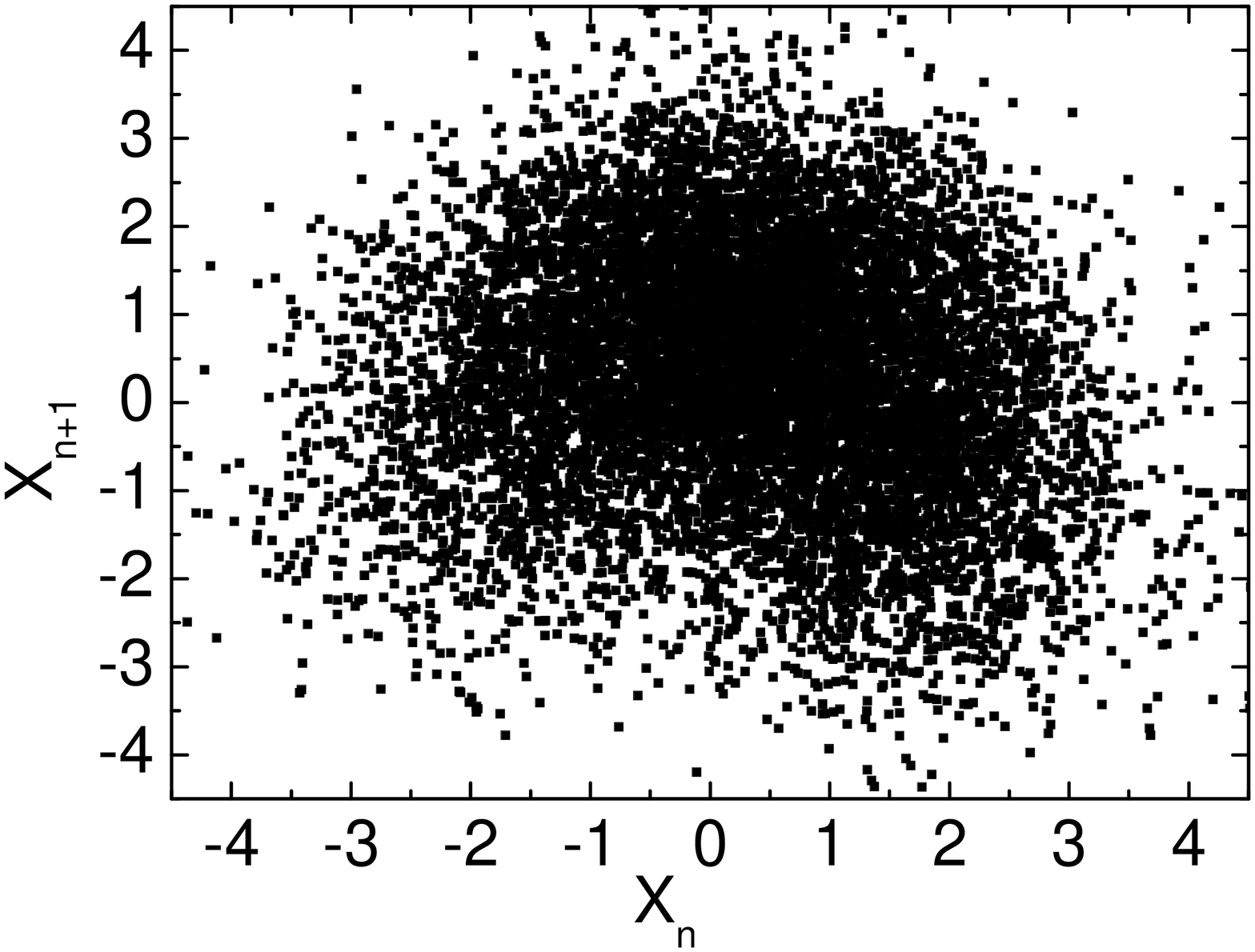}
\caption{\label{fig.henon100b} Chaotic Henon map with a
measurement noise $\mathcal{N}=69\%$. Note the difference in
scale.}
\end{figure}
\begin{figure}
\includegraphics[angle=-90,scale=0.35]{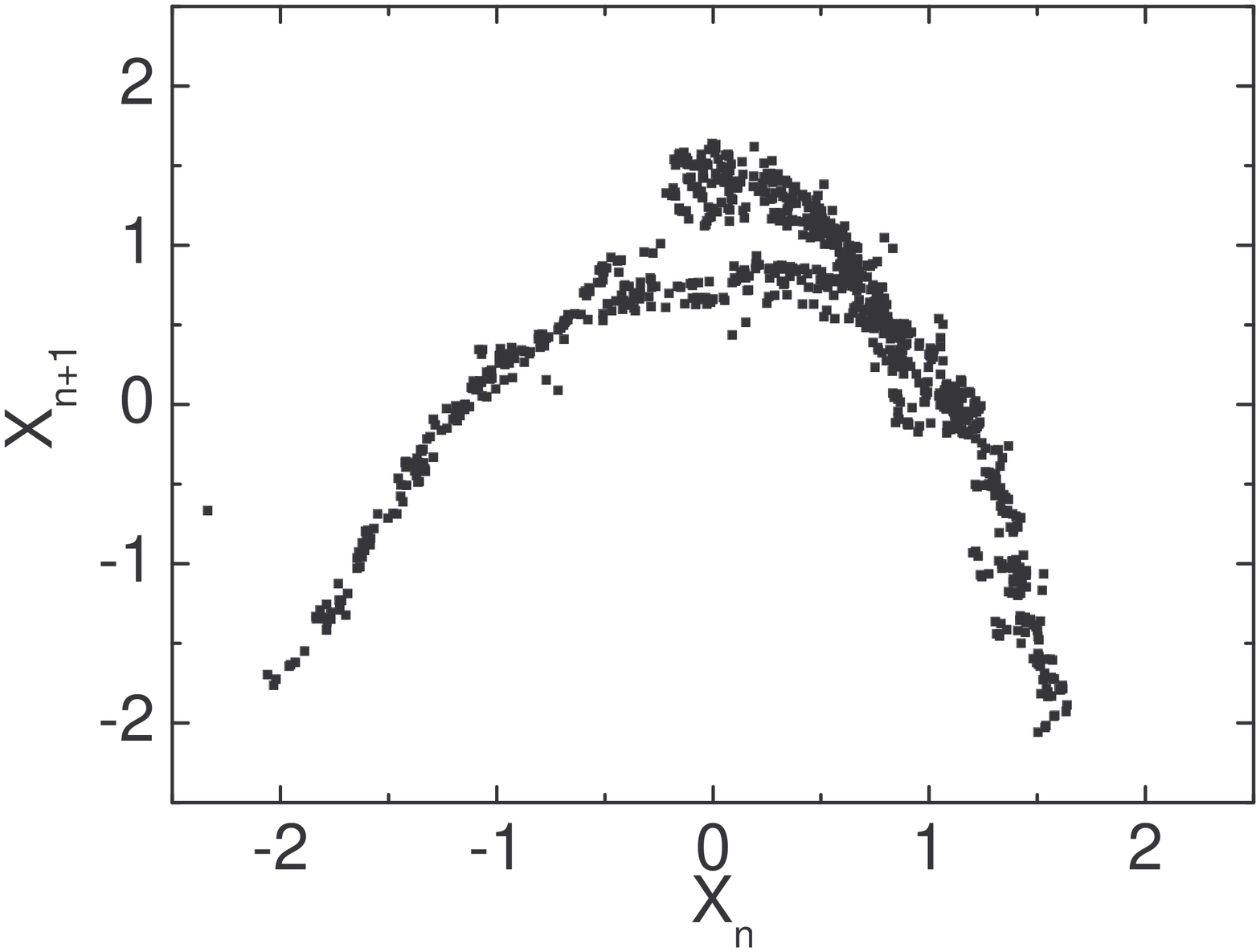}
\caption{\label{fig.henon100c} Chaotic Henon map with a
measurement noise after noise reduction with the LPNC method,
$\mathcal{G}=4.3$ }
\end{figure}

The LPNC method has been applied to three systems: the Henon map,
the Lorenz model \cite{lorenz} and the Chua circuit
\cite{anischenko,chua1,chua2}. Figures \ref{fig.henon100a} -
\ref{fig.henon100c} present the chaotic Henon map in the absence
and in the presence of measurement noise as well as a result of
the noise reduction.
Table~\ref{tab.smallnoise} presents the values of the gain
parameter for the Henon map and for the Lorenz system.

\begin{figure}
\includegraphics[angle=-90,scale=0.35]{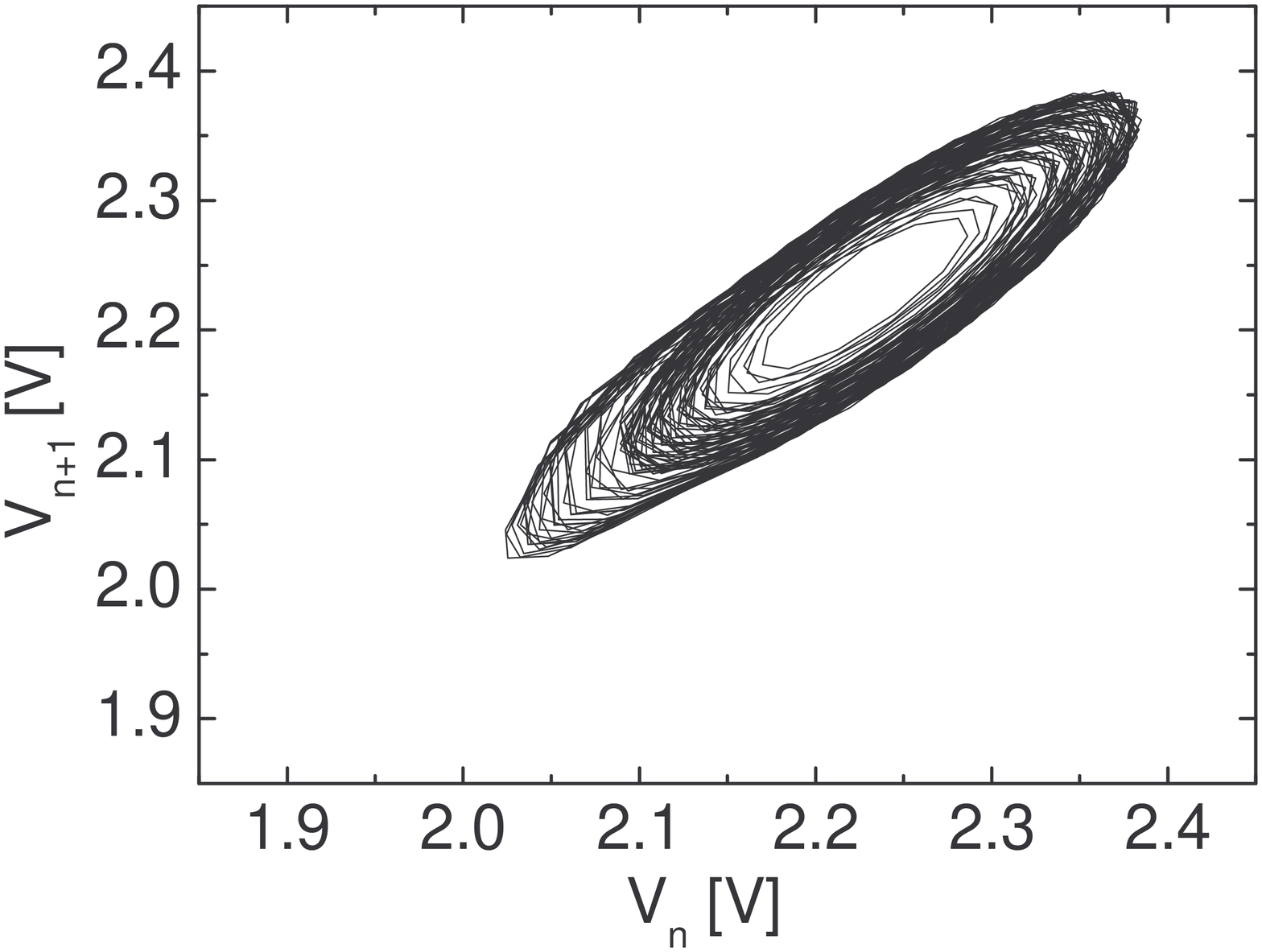}
\caption{\label{fig.chua7000} The stroboscopic map corresponding
to a clean trajectory in the Chua circuit.}
\end{figure}

\begin{figure}
\includegraphics[angle=-90,scale=0.35]{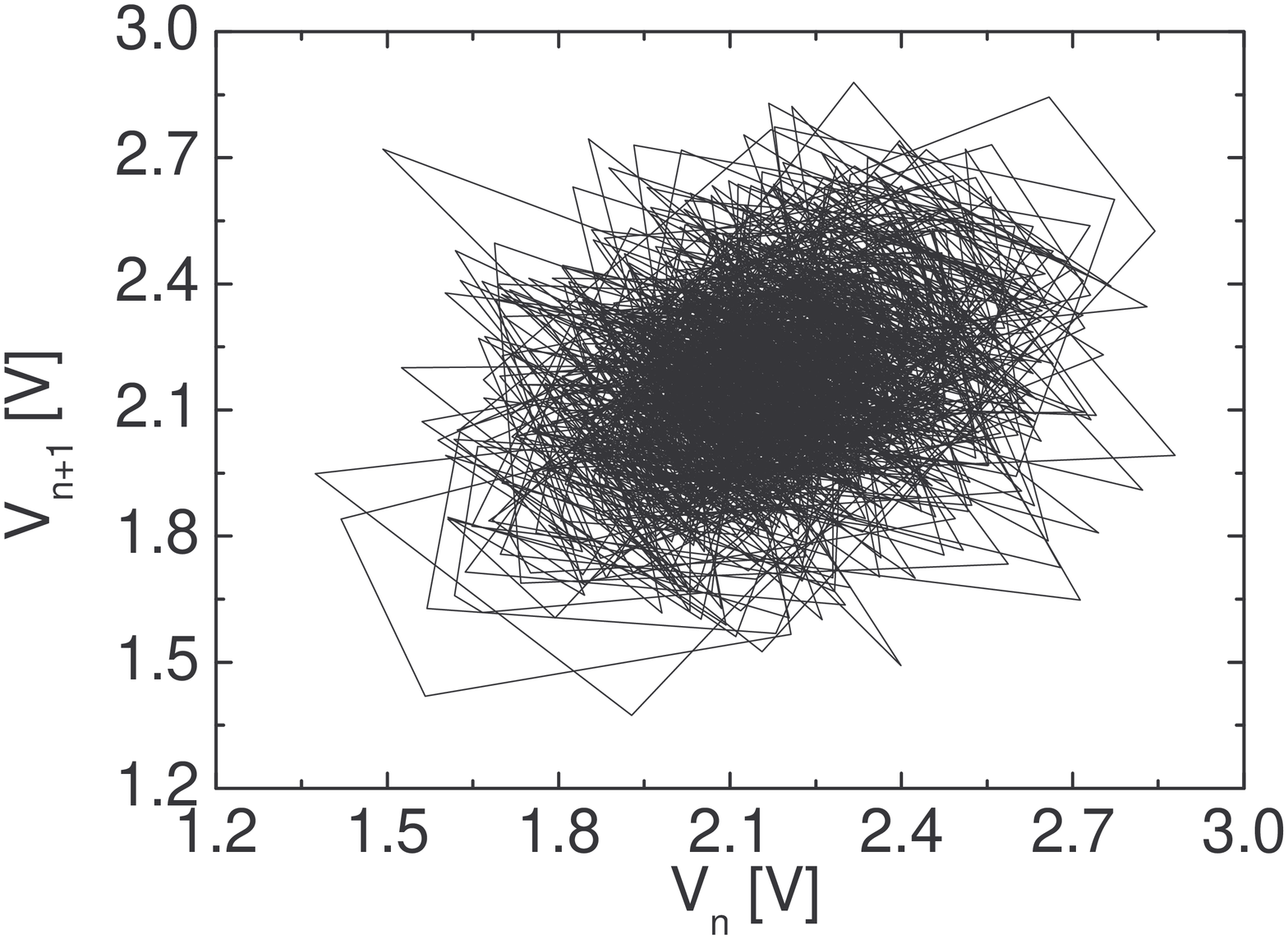}
\caption{\label{fig.chua7000no} The stroboscopic map received from
the Chua circuit in the presence of a measurement noise
$\mathcal{N}=96.5\%$. Note the difference in scale.}
\end{figure}

\begin{figure}
\includegraphics[angle=-90,scale=0.35]{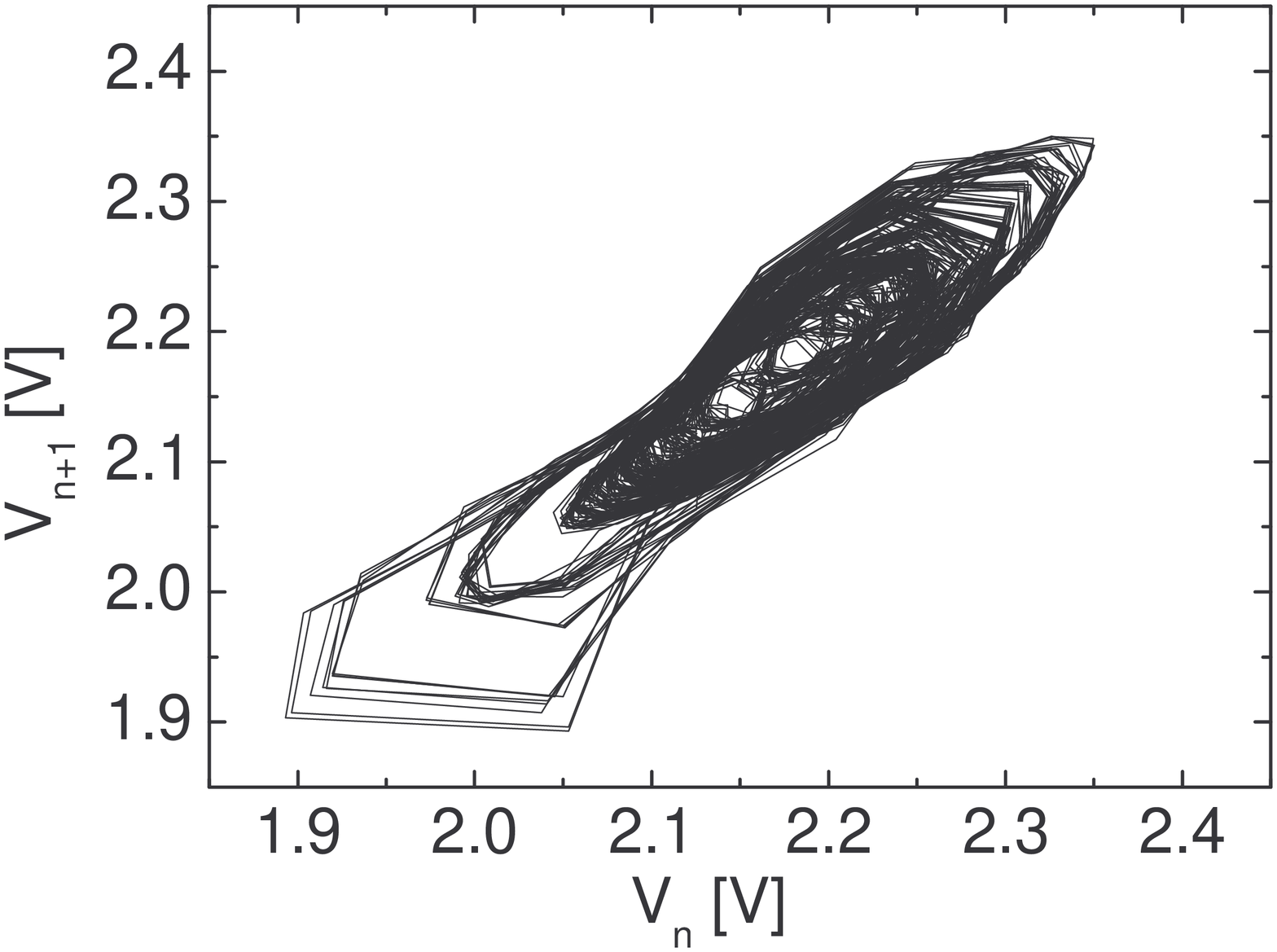}
\caption{\label{fig.chua7000cl} The stroboscopic map received
after the noise reduction by the LPNC method applied to data
presented at Fig.~\ref{fig.chua7000no}.}
\end{figure}

To verify our method in a real experiment we have performed the
analysis of data generated by a nonlinear electronic circuit. The
Chua circuit in the chaotic regime \cite{chua1,chua2} has been
used and we have added a measurement noise to the outcoming
signal. The noise (white and Gaussian) came from an electronic
noise generator. Figures \ref{fig.chua7000} - \ref{fig.chua7000cl}
show a clean signal coming from this circuit, the signal generated
by Chua circuit with measurement noise ($\mathcal{N}=96.5\%$) and
the same signal after the noise reduction with the LPNC method
($\mathcal{G}=6.38$). Table~\ref{tab.red2} presents values of the
$\mathcal{G}$ parameter and the percentage of eliminated noise for
several values of the noise level in the Chua circuit.

\begin{table}
\caption{\label{tab.smallnoise} Results of noise reduction by the
LPNC method for the Henon map and Lorenz model.}
\begin{ruledtabular}

\begin{tabular}{cccc}

 System & $\mathcal{N}$&  $\mathcal{G}$& percent. of eliminated noise\\
\hline Henon $N=1000$ &$10\%$ &$9.58$ &$89\%$ \\
Henon $N=3000$& $10\%$& $10.04$ & $91\%$ \\
Henon $N=1000$& $66\%$& $5.08$ & $69\%$\\
Lorenz $N=1000$& $78\%$& $5.85$ & $74\%$\\
Lorenz $N=3000$ & $76\%$& $6.02$ & $75\%$\\
Lorenz $N=1000$& $34\%$& $7.21$ & $81\%$\\
\end{tabular}
\end{ruledtabular}
\end{table}

\begin{table}
\caption{\label{tab.red2} Results of noise reduction by the LPNC
method for the Chua circuit with a measurement noise ($N=3000$).}
\begin{ruledtabular}

\begin{tabular}{ccc}

 $\mathcal{N}$&  $\mathcal{G}$& percent. of eliminated noise\\
\hline
   $24.9\%$& 5.4& 71\%\\
   $28.3\%$& 4.9&68\%\\
$46.1\%$ &  7.0 & 80\% \\
$73.7\%$ & 4.81 & 67\% \\
$ 90.6\%$ & 7.4 & 82\% \\
$96.5\%$ & 6.4 &  77\% \\

\end{tabular}
\end{ruledtabular}
\end{table}

\begin{figure}
\includegraphics[angle=-90,scale=0.35]{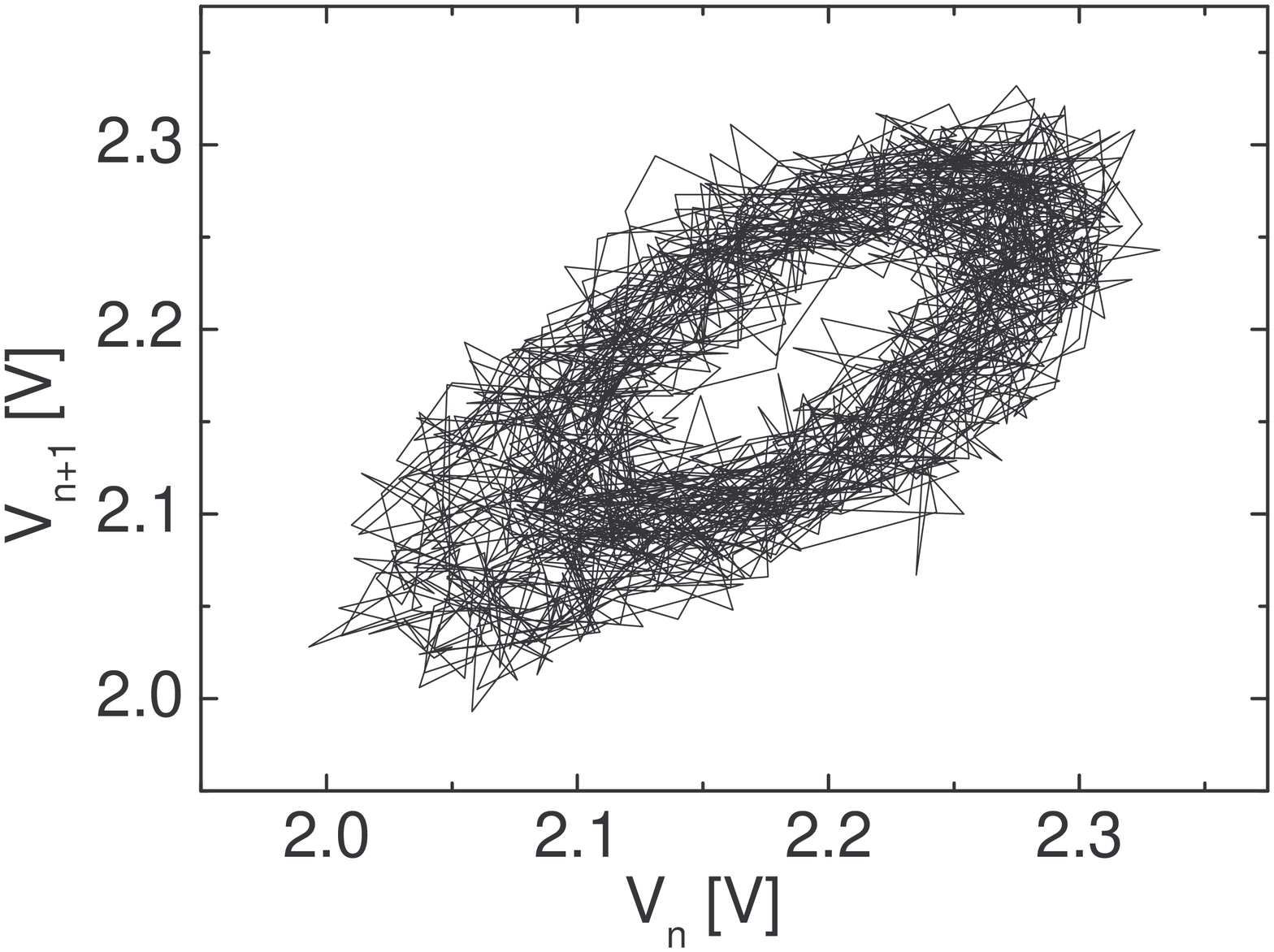}
\caption{\label{fig.chuadnno} Stroboscopic map received from the
Chua circuit in the presence of a mixture of a measurement and
dynamical noise $\mathcal{N}\approx 22\%$.}
\end{figure}

\begin{figure}
\includegraphics[angle=-90,scale=0.35]{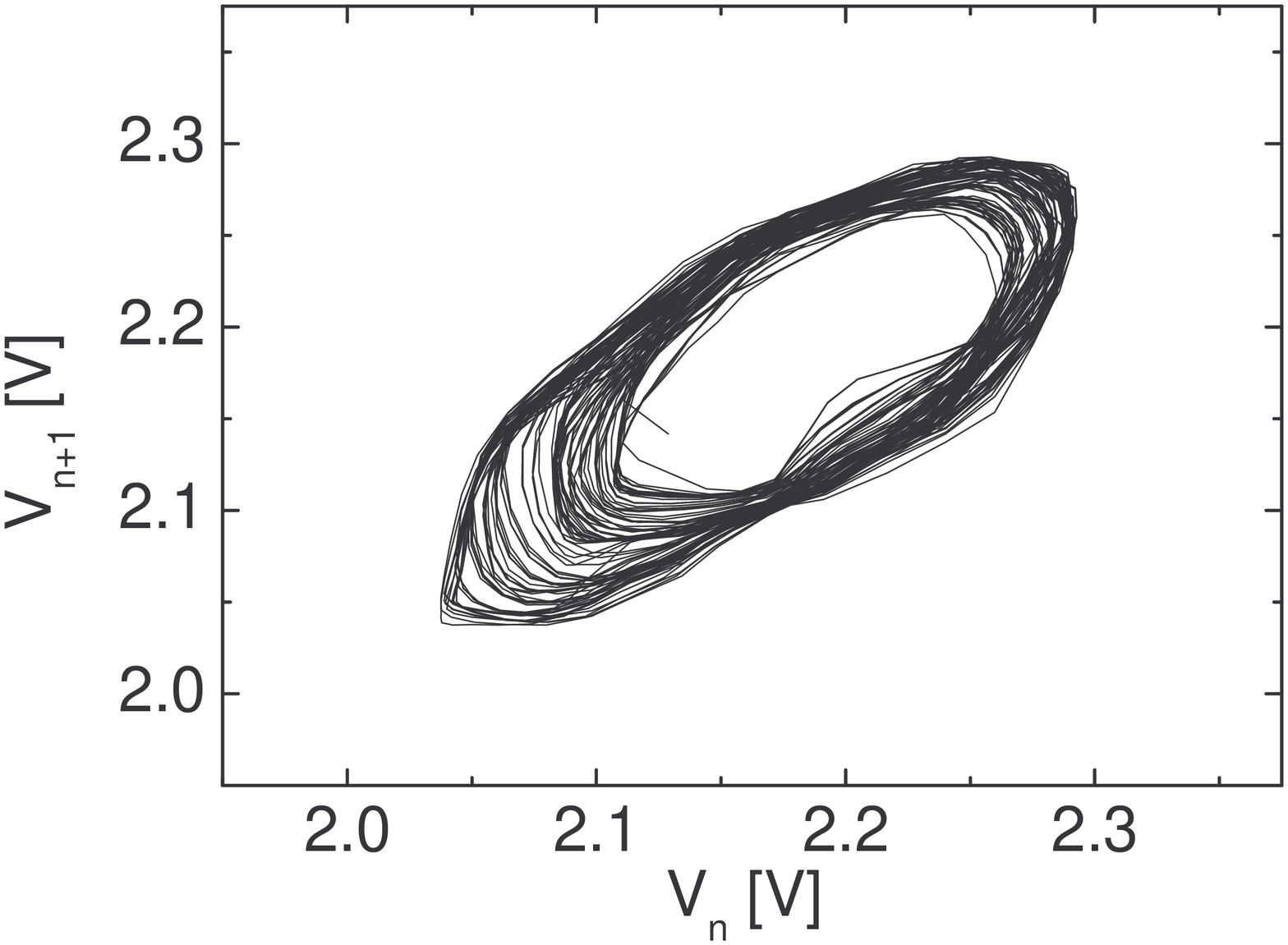}
\caption{\label{fig.chuadncl} Stroboscopic map received after
noise reduction by the LPNC method applied to data presented at
Fig.~\ref{fig.chuadnno}}
\end{figure}

 All the above applications of the LPNC method consider the case of measurement
 noise that has been added to the signal in numerical or electronic experiments.
  However, our LPNC method can also be applied to dynamical noise
i.e to the noise which in experiments is included in the equations
of motion \cite{Jaeger}. In such a case one cannot compare the
noisy data with the clean trajectory since the latter one does not
exist anymore, and there are only $\epsilon$-shadowed trajectories
\cite{Farmer} that can be approximated by means of the LPNC
method. Figure~\ref{fig.chuadnno} shows a measured signal
generated by a Chua circuit where a mixture of measurement noise
and dynamical noise occurs. Figure~\ref{fig.chuadncl} shows the
result of noise reduction applied to such a signal.

\section{\label{sec.estimation}Noise level estimation by LPNC method}
\par The LPNC method introduced in the previous section can be used to quantify the noise
level of data. The noise level, i.e. the standard deviation in
noisy time series, may be approximated as the Euclidian distance
between the vectors $\set{x_i}$ and $\set{\bar{x}_i}$ representing
the time series before and  after noise reduction \cite{Hsu} \be
\tilde{\sigma}_{noise}\approx\sqrt{\frac{1}{N}\sum\limits_{i=1}^N
\lb x_i-\bar{x}_i\rb^2}.\label{eq.oszac} \end{equation} The main
disadvantage of the LPNC method used for the noise level
estimation is its small rate of convergence with respect to other
known methods \cite{urbanowicz,Diks,Yu,Oltmans} and the fact that
the method can be used only for low-dimensional systems. On the
other hand the LPNC method can be applied for estimation of any
noise level including a large one.  In Table~\ref{tab:tab.1} we
have presented the estimated noise level $\tilde{\sigma}_{noise}$
for the Chua circuit.
\begin{table}
\caption{\label{tab:tab.1} Noise level estimated by the LPNC
method for the Chua circuit with measurement noise ($N=3000$).}
\begin{ruledtabular}

\begin{tabular}{ccc}

 $\mathcal{N}$& $\sigma_{noise}$ [mV]& $\tilde{\sigma}_{noise}$ [mV]\\
\hline
 $0\%$& 0& 5.5 \\
 $3.1\%$&30.4& 28.9\\
  $6.2\%$&   60.8& 53.7\\
  $12.3\%$&  121.7&110 \\
   $24.9\%$& 243.4& 235\\
   $28.3\%$& 304&305\\
$46.1\%$ &   486& 454 \\
$73.7\%$ & 973 & 938 \\
$ 90.6\%$ & 1520 & 1375 \\
$96.5\%$ & 2120 &  1844 \\

\end{tabular}
\end{ruledtabular}
\end{table}

\section{\label{sec.conclusions}Conclusions}
\par In conclusion we have developed a method of noise reduction
that makes use of nonlinear constraints which occur in a natural
way due to the linearization of a deterministic system trajectory
in the nearest neighborhood of every point in the phase space.
This neighborhood has been determined by Delaunay triangulation.
The method has been applied to data from the Henon map, Lorenz
model and electronic Chua circuit contaminated by measurement
(additive) noise. The efficiency of our method is comparable to
that of standard LP methods but it is more robust to input
parameter adjustment.

\begin{acknowledgments}
We grateful acknowledge helpful discussion  with Holger Kantz and
Rainer Hegger. KU is thankful to Organizers of the Summer School
\textit{German-Polish Dialogue 2002} in Darmstadt. He has
partially been supported by the KBN Grant 2 P03B 032 24 and JAH
has been supported by the special program \textit{Dynamics of
Complex Systems} of Warsaw University of Technology.
\end{acknowledgments}

\newpage
\par
\appendix
\section{\label{sec.multidimensional} Multidimensional version of LPNC method}
\par In Sec.~\ref{sec.1dimLPC} the LPNC method has been presented for one-dimensional systems.
Here we show the generalization of this approach for
$d$-dimensional dynamics. For one-dimensional problems the Jacobi
matrix of the system does not appear explicitly in our method. For
higher dimensional models the corresponding Jacobian $\mathbf{A}$
has to be calculated but we manage to minimalize  errors occurring
by its estimation. The linearized equation of motion for vectors
from the nearest neighborhood $\tNN$ of a vector
$\tilde{\mathbf{x}}_n$ can be written in the form \be
\tilde{\mathbf{x}}_{n+1}=\mathbf{A}\cdot
\tilde{\mathbf{x}}_n+\mathbf{b}.\label{eq.zlin2D}\end{equation} In
such a case one needs three vectors
$\tilde{\mathbf{x}}_n,\tilde{\mathbf{x}}_k,\tilde{\mathbf{x}}_j
\in \tNN$ to write constraints corresponding to
Eq.~(\ref{eq.wiaz1D}). In comparison with to the one-dimensional
case the number of near neighbors i.e. the number of points in the
set $\tNN$ must be larger to allow a unique estimation of the
Jacobian $\mathbf{A}$. We assume that the Jacobi matrix can be
approximately received by minimalization of the following cost
functional
\begin{eqnarray} \sum_s (
a_{m1}x_s+a_{m2}x_{s-\tau}+\ldots\nonumber\\\ldots+a_{md}x_{s-\lb
d-1\rb\tau}-x_{s+1-m\tau})^2= min\nonumber\\
(\forall_{m=1,\ldots,d})\quad\mbox{and}\quad\set{s:\mathbf{x}_s\in\NN},\end{eqnarray}
where $a_{ij}=\left[\mathbf{A}\right]_{ij}$. By analogy with
Eq.~(\ref{eq.wiaz1D}) we introduce the follow

\begin{eqnarray}
   G_m^d \lb \tNN\rb \equiv a_{m1}G\lb
   \tNN,\mathbf{\tilde{X}}_{n+1-m\tau}\rb+\nonumber\\a_{m2}G\lb
   \mathbf{\tilde{X}}_{n-\tau},\mathbf{\tilde{X}}_{n+1-m\tau}\rb+\ldots\nonumber\\
   \ldots+a_{md}G\lb
   \mathbf{\tilde{X}}_{n-\lb
   d-1\rb\tau},\mathbf{\tilde{X}}_{n+1-m\tau}\rb=0\nonumber\\\quad(\forall_{m=1,\ldots,d}),\label{eq.mulwiaz2D}
\end{eqnarray}
where we used the notation corresponding to
equation~(\ref{eq.rozwin}) i.e.
\begin{widetext}\begin{eqnarray} G\lb \NN,
\mathbf{X}_{n+l}\rb=x_n\lb x_{k+l}-x_{j+l}\rb+x_k\lb
x_{j+l}-x_{n+l}\rb+x_j\lb
x_{n+l}-x_{k+l}\rb\nonumber\\
G\lb \mathbf{X}_{n-s}, \mathbf{X}_{n+l}\rb=x_{n-s}\lb
x_{k+l}-x_{j+l}\rb+x_{k-s}\lb x_{j+l}-x_{n+l}\rb+x_{j-s}\lb
x_{n+l}-x_{k+l}\rb,
\end{eqnarray}\end{widetext}
$\mathbf{X}_{n+s}=\set{\mathbf{x}_{n+s},\mathbf{x}_{k+s},\mathbf{x}_{j+s}}\quad(\forall_{s=0,\pm
1,\pm 2,\ldots})$ where
$\set{n,j,k:\mathbf{x}_{n},\mathbf{x}_{k},\mathbf{x}_{j}\in\NN}$.
Since the clean trajectory is not known thus in the
Eq.~(\ref{eq.mulwiaz2D}) the observed variables $\NN$,
$\mathbf{X}_{n+1-m\tau}$  etc. are used.

In such a way the equation~(\ref{eq.przybliz}) can be written in a
more general way as \begin{eqnarray} \sum_{l=1}^d a_{ml}G\lb\delta
\mathbf{X}_{n-\lb l-1\rb\tau}, \delta \mathbf{X}_{n+1-m\tau}\rb
\cong
0\nonumber\\\quad(\forall_{n=1,\ldots,N},\forall_{m=1,\ldots,d})\label{eq.przybliz2D}\end{eqnarray}
where we use
\begin{eqnarray} G\lb\delta \mathbf{X}_{n-s},\delta
\mathbf{X}_{n+l}\rb=\delta x_{n-s}\lb \delta x_{k+l}-\delta
x_{j+l}\rb\nonumber\\+\delta x_{k-s}\lb \delta x_{j+l}-\delta
x_{n+l}\rb+\delta x_{j-s}\lb \delta x_{n+l}-\delta
x_{k+l}\rb\end{eqnarray} $\delta \mathbf{X}_{n+s}=\set{\delta
\mathbf{x}_{n+s},\delta \mathbf{x}_{k+s},\delta
\mathbf{x}_{j+s}}\quad(\forall_{s=0,\pm 1,\pm 2,\ldots})$, where
$\set{n,j,k:\mathbf{x}_{n},\mathbf{x}_{k},\mathbf{x}_{j}\in\NN}$,
$\delta \mathbf{x}_n=\set{\delta x_n,\delta
x_{n-\tau},\ldots,\delta x_{n-\lb d-1\rb\tau}}\quad\mbox{and}\quad
\set{n:\mathbf{x}_n\in\NN}$.
\par Now the cost problem~(\ref{eq.minimalizlok}) can be transformed to the form
    \begin{eqnarray}
    \tilde {S}_n^{NN}=\sum_s \lb \delta x_s
    \rb^2+\lambda_n^m G_m^d \lb \NN\rb = min\nonumber\\ (\forall_{n=1,...,N},\forall_{m=1,\ldots,d})
    \nonumber\\\quad\mbox{and} \quad\set{s:\mathbf{x}_s\in \NN\;\mbox{or}\; \mathbf{x}_s\in
    \mathbf{X}_{n+1}}.\label{eq.minimalizlok2D}
    \end{eqnarray}
Finding zeros of partial derivatives of the functional
(\ref{eq.minimalizlok2D}) one can linearize this problem and write
it in the form similar to the Eq.~(\ref{eq.liniowezlok}) \be
 \mathbf{M}_n\cdot\delta
\mathbf{X}_n^{\lambda}=\mathbf{B}_n\quad(\forall_{n=1,\ldots,N}).\label{eq.linlok2D}\end{equation}
Vectors $\delta \mathbf{X}_n^{\lambda}$ and $\mathbf{B}_n$
occurring in Eq.~(\ref{eq.linlok2D}) are equal to
\begin{eqnarray}\lb\delta \mathbf{X}_n^{\lambda}\rb^T=\{\delta
x_{n-\lb d-1\rb\tau},\delta x_{n-\lb
d-2\rb\tau},\ldots\nonumber\\\ldots,\delta x_{n},\delta
x_{n+1},\lambda_n^1,\lambda_n^2,\ldots,\lambda_n^d\},\end{eqnarray}
\begin{eqnarray}\mathbf{B}_n^T=\{0,0,\ldots,0,-G_1^d\lb\NN\rb,-G_2^d\lb\NN\rb,\ldots\nonumber\\
\ldots,-G_d^d\lb\NN\rb\}\end{eqnarray} where the number of zeros
$d_0$ appearing in $\mathbf{B}_n^T$ depends on the values of
$\tau$ and $d$ and for the case $\tau=1$, $d_0=d+1$. Elements of
the matrix $\mathbf{M}_n$ can be written as
\begin{widetext}
\begin{eqnarray}
\left[\mathbf{M}_n\right]_{mm}&=&2\quad(\forall_{m=1,\ldots,d_0})\nonumber\\
 \left[\mathbf{M}_n\right]_{lm}&=&\left[\mathbf{M}_n\right]_{ml}+=\sum_{s=1}^d
a_{ms} \lb x_{k+1}-x_{j+1}\rb
\quad(\forall_{m=d_0+1,\ldots,d_0+d+1})\quad \mbox{and} \quad
\set{l:x_l\in\mathbf{x}_n}\nonumber\\
\left[\mathbf{M}_n\right]_{lm}&=&\left[\mathbf{M}_n\right]_{ml}+=\sum_{s=1}^d
a_{ms} \lb x_{j+1}-x_{n+1}\rb
\quad(\forall_{m=d_0+1,\ldots,d_0+d+1})\quad \mbox{and} \quad
\set{l:x_l\in\mathbf{x}_k}\nonumber\\
\left[\mathbf{M}_n\right]_{lm}&=&\left[\mathbf{M}_n\right]_{ml}+=\sum_{s=1}^d
a_{ms} \lb x_{n+1}-x_{k+1}\rb
\quad(\forall_{m=d_0+1,\ldots,d_0+d+1})\quad \mbox{and} \quad
\set{l:x_l\in\mathbf{x}_j}\label{eq.wielkierown}\\
\left[\mathbf{M}_n\right]_{lm}&=&\left[\mathbf{M}_n\right]_{ml}+=\sum_{s=1}^d
a_{ms} \lb x_{j-m\tau}-x_{k-m\tau}\rb
\quad(\forall_{m=d_0+1,\ldots,d_0+d+1})\quad \mbox{and} \quad
\set{l:x_l\in\mathbf{x}_{n+1}}\nonumber\\
\left[\mathbf{M}_n\right]_{lm}&=&\left[\mathbf{M}_n\right]_{ml}+=\sum_{s=1}^d
a_{ms} \lb x_{n-m\tau}-x_{j-m\tau}\rb
\quad(\forall_{m=d_0+1,\ldots,d_0+d+1})\quad \mbox{and} \quad
\set{l:x_l\in\mathbf{x}_{k+1}}\nonumber\\
\left[\mathbf{M}_n\right]_{lm}&=&\left[\mathbf{M}_n\right]_{ml}+=\sum_{s=1}^d
a_{ms} \lb x_{k-m\tau}-x_{n-m\tau}\rb
\quad(\forall_{m=d_0+1,\ldots,d_0+d+1})\quad \mbox{and} \quad
\set{l:x_l\in\mathbf{x}_{j+1}}\nonumber
\end{eqnarray}\end{widetext}
where the remaining $\mathbf{M}_n$ elements vanish and $x_l\in
\mathbf{x}_n$ means that the variable $x_l$ is a component of the
$\mathbf{x}_n$ vector.
\par The operator $+=$ in (\ref{eq.wielkierown}) has the same meaning as in the programming language
C++, i.e. if elements of the matrix $\left[\mathbf{M}_n\right]_{
ml }$ occur in a few places (e.g. : $ x_n\in
\mathbf{x}_n\;\mbox{and}\; x_n\in\mathbf{x}_{n+1}\quad
\forall_{d>1},\vee_{\tau=1}$) then the elements at the rhs of such
equations have to be summed up.

\end{document}